\documentclass[twocolumn]{aastex62}
\usepackage[normalem]{ulem}
\usepackage{graphicx}
\usepackage{xcolor}
\usepackage{amsmath,mathrsfs}
\usepackage{amssymb} 
\usepackage{soul}

\newcommand{\ms}{M$_{\odot}$}
\newcommand{\eps}{{\large{$\epsilon$}\ }}

\newcommand{\co}{$^{12}$CO}
\newcommand{\coc}{$^{13}$CO}
\newcommand{\coo}{C$^{18}$O}

\received{January 1, 2021}
\revised{January 7, 2021}
\accepted{November 8, 2021}
\submitjournal{ApJ}

\shorttitle{Disk gas masses from C$^{18}$O emission} 

\begin{document}
\title{C$^{18}$O emission as an effective measure of gas masses of protoplanetary disks}

\correspondingauthor{M. Ruaud}
\email{maxime.ruaud@gmail.com}

\author{Maxime Ruaud} 
\affiliation{NASA Ames Research Center, Moffett Field, CA, USA}
\affiliation{Carl Sagan Center, SETI Institute, Mountain View, CA, USA} 

\author{Uma Gorti}
\affiliation{NASA Ames Research Center, Moffett Field, CA, USA}
\affiliation{Carl Sagan Center, SETI Institute, Mountain View, CA, USA} 

\author{David J. Hollenbach}
\affiliation{Carl Sagan Center, SETI Institute, Mountain View, CA, USA} 


\begin{abstract}
Many astrochemical models of observed CO isotopologue line emission, earlier considered a good proxy measure of H$_2$ and hence disk gas mass, favor large deviations in the carbon and oxygen gas phase abundances and argue that severe gas phase CO depletion makes it a poor mass tracer. Here, we show that C$^{18}$O line emission is an effective measure of the gas mass, and despite its complex chemistry, a possibly better tracer than HD. Our models are able to reproduce C$^{18}$O emission from recent ALMA surveys and the TW Hya disk to within a factor of $\sim 2-3$ using carbon and oxygen abundances characteristic of the interstellar medium (C/H$=1.4 \times 10^{-4}$; O/H$=3.2\times 10^{-4}$) without having to invoke unusual chemical processing. Our gas and dust disk structure calculations considering hydrostatic pressure equilibrium and our treatment of the CO conversion on grains are primarily responsible for the very different conclusions on disk masses and CO depletion. As did previous studies, we find that a gas phase C/O of $\sim 1-2$ can explain observed hydrocarbon emission from the TW Hya disk; but significantly, we find that CO isotopologue emission is only marginally  affected by the C/O ratio. We therefore conclude that C$^{18}$O emission provides estimates of disk masses that are uncertain only to within a factor of a few, and describe a simplified modeling procedure to obtain gas disk masses from C$^{18}$O emission lines.
\end{abstract}

\section{Introduction} \label{sec:intro}
 
 The question of how much mass is present in protoplanetary disks is fundamental to understanding disk evolution and planet formation. The rate at which the total mass declines with time is the primary indicator of the efficiency of mass and angular momentum transport, and of planet formation and disk dispersal timescales. The available reservoirs of gas and dust masses, how they evolve with time and relative to one another, as well as how the disk chemical composition itself changes are all relevant to determining the conditions under which rocky and gaseous planets can form. 
 
Dust masses for particles $\lesssim$ 1 cm in size can be determined from submillimeter continuum emission to within an order of magnitude  despite some uncertainties in opacity, as dust emission from particles with sizes $\lesssim$ 1 cm is mostly optically thin with the emission proportional to the dust mass \citep[e.g.][]{Hildebrand1983, Ansdell16,Pascucci16, Tychoniec2020}. 
 
 Gas mass determination, however, has been difficult even to order of magnitude certainty as there are few optically thin tracers of gas.  HD, which is the isotopologue of the main gas constituent H$_2$, is expected to be optically thin for typical disks but has been detected only in three disks by Herschel \citep{Bergin13,Mcclure2016}. Both the lack of facilities sensitive enough to detect the faint line \citep[e.g.][]{Bergin18} and the fact that disk regions containing most of the gas mass are still too cold to excite the $E_u \sim 128$K upper level of HD \citep[e.g.][]{Trapman17} limit our ability to use this molecule as a gas mass tracer. CO is the next most abundant molecule and its isotopologues are often used as tracer species; several surveys of disks in star-forming regions have been now conducted with ALMA measuring both dust continuum emission and \coo\ line emission \citep{Ansdell16,Ansdell17,Long17}. However, most recent studies  conclude that \coo \  may not be a reliable indicator of gas disk mass \citep[e.g.][]{Calahan21} as ascribed to reasons below.

CO isotopologue emission from disks has been extensively investigated in recent years.  CO emission is bright, and with many accessible transitions and isotopologues (e.g., \coc, \coo), can potentially trace a wide range of temperatures and densities. CO emission can also be spatially mapped using interferometers such as ALMA and, in principle, a complete survey of CO and its isotopologues can be used to retrieve and piece together the disk radial gas surface density distribution \citep[e.g., as has been done for TW Hya or GM Aur, see][]{Zhang17,Zhang19,Calahan21,Schwarz21}. However, CO can freeze out on the surface of cold dust grains, and characteristic dust temperatures in the outer disk midplanes are such that these regions are obscured due to a lack of gas phase CO. Even after accounting for the fact that CO condenses in the midplane (at $T_d\sim 20$K, assuming a CO binding energy of 1150K and no further chemical processing in the ice), models have typically overproduced \coo\ fluxes leading to speculations about a variable CO/H$_2$ ratio in disks. However, a key issue is that CO emission does not probe total disk mass directly---optically thin CO isotopologues are photodissociated at the surface and CO freezes out near the midplane---and models are needed to correlate the observed flux with the emission; the conclusions made are thus dependent on the model assumptions. Rarer isotopologues involve multiplications by larger factors to convert to H$_2$, and even minor uncertainties in their isotopic fractionation chemistry can get amplified to make such gas mass determinations unreliable. 

Recent ALMA suveys have detected line fluxes for CO and its isotopologues and these are far lower than many model predictions that assume canonical ISM CO/H$_2$ and dust/gas mass ratios \citep{Ansdell16,Miotello17,Long17}. These low fluxes are usually interpreted using two scenarios: (1) either that CO is a good measure of gas mass and there is a significant gas removal  leading to low gas masses in these disks, (2) or that CO is severely under-abundant throughout these disks (i.e. not only in the disk midplane where CO freezes-out) and that gas phase CO is depleted even at the surface by some efficient mechanism. For (1), \citet{Ansdell16}, using the simple model for disk structure and chemistry from \citet{Williams14} (developed to interpret SMA data of a survey of 15 disks), conclude that most observed disks in Lupus have low gas masses, with dust/gas mass ratios $\sim 1$ which implies a factor of 100 reduction from that in the parent cloud material, indicating relatively rapid disk gas dispersal ; which however is at odds with the relatively high mass accretion rates inferred in the region \citep{Manara16,Alcala17}. These were revised downwards by a factor of  $\sim 10$ using more sophisticated disk models that include selective photodissociation of \coo\ \citep{Miotello14,Miotello16,Miotello17}. However, Herschel observations of HD $(1-0)$ in a few disks find dust/gas mass ratios that are compatible with ISM values \citep{Bergin13,Mcclure2016}. These findings have led many authors to favor (2) and therefore invoke large ($\sim 2$ orders of magnitude) depletion of CO that must act in addition to processes like CO freeze-out in the cold midplane \citep[e.g.][]{Favre13,Du15,Schwarz16,Kama16,Trapman17}. This deficit in observed CO isotopologue emission (compared to observed HD emission and model CO predictions) has been explained as being due to a change in C/H and C/O ratios in disks as they evolve \citep[e.g.][]{Du15,Bergin16,Kama16,Schwarz16,Xu17,Krijt16,Krijt18,Krijt20,Zhang19,McClure19,Calahan21,Zhang21,Schwarz21} or as the result of the incorporation of CO into solids on the form of more complex compounds at the surface of the grains such as CO$_2$, CH$_3$OH and carbon chains \citep[e.g.][]{Bergin14,Furuya14,Aikawa15,Reboussin15,Molyarova17,Eistrup16,Bosman18,Ruaud19}. However, the more widely held interpretation is that pebble formation in disks locks up available carbon and oxygen and gradually depletes gas phase CO \citep{Krijt18,Krijt20}. This is then believed to change the global C/O and C/H ratios in disks, lowering the CO gas phase abundance in emitting regions from canonical ISM values. A more complete discussion of the chemistry involved and the implications for gas mass is presented in a recent review by \citet{Oberg21}. 

In this paper, we examine the global CO isotopologue depletion issue in greater detail and argue against the need for both scenarios (1) and (2). In order to determine the extent to which CO isotopologue emission can measure disk mass, one needs to determine both the extent to which C/O changes and to which C and O abundances need to be altered to produce the observed emission. We revisit the problem with newer, more complete models of disk chemistry \citep{Ruaud19}. Our main motivation for this study is that none of the previous models have simultaneously considered all the processes critical to interpreting CO emission (i) a full treatment of grain chemistry and CO freeze-out, (ii) CO isotopologue selective photodissociation, and (iii) a self-consistent gas disk density and temperature structure coupled by vertical pressure equilibrium. Using models that incorporate all these physical and chemical processes, we conduct a parameter survey where we vary gas disk masses and the dust/gas mass ratio. We investigate the effect of grain surface chemistry on CO abundance, the dependence of CO emission on disk parameters, and also evaluate other mass tracers like HD and attempt to reconcile model results with available observational data. We show that C$^{18}$O emission is well explained with disks having having ISM like elemental abundances and argue that \coo\ is a good tracer of disk gas mass over a wide range of disk masses. We also show that observed CO emission does not by itself indicate any change in the C/O ratio in disks, and discuss the key model assumptions necessary to retrieve disk masses from CO isotopologue emission. The structure of the paper is as follows. After a brief description of the models in \S 2, the main results are presented in \S 3, followed by a comparison with existing models and data in \S 4 and prescriptions for simple analyses in \S 5. Conclusions are presented in \S 6. 

\section{Model description}
\label{sec:model_desc}
We use the modeling framework described in \citet{Ruaud19} (hereafter RG19) to compute the disk physical structure and chemistry. A brief summary of the modeling procedure is provided here and we refer the reader to RG19 for further details. Stellar properties (mass, spectrum, UV and X-ray spectra) and the disk gas surface density ($\Sigma_\mathrm{gas}(r)$) and dust surface density ($\Sigma_\mathrm{dust}(r)$) are the input parameters. The disk physical structure is determined by solving for vertical hydrostatic gas pressure equilibrium simultaneously with a surface density constraint; 
\begin{equation}
    \frac{dP(r,z)}{dz} = -\rho(r,z) \Omega^2 z; \quad \Sigma_\mathrm{gas}(r) = \int \rho(r,z) dz
    \label{eq:vhse} 
\end{equation}
where the pressure $P(r,z) = \rho(r,z) k_B T(r,z)/\mu(r,z)$ and  $\rho(r,z) , T(r,z), \mu(r,z)$ and $\Omega$ are the gas density, temperature, mean molecular weight and Keplerian frequency respectively. Gas heating includes dust collisions, UV, X-rays and other mechanisms, cooling is by line and continuum emission. 
At each $(r,z)$ the gas density, temperature and chemistry and the dust size distribution are all coupled and we find a consistent solution that matches the local pressure gradient and results in the specified $\Sigma_\mathrm{gas}(r)$ and $\Sigma_\mathrm{dust}(r)$. At a given radius $r$, the dust grain size distribution is determined assuming a balance between fragmentation and coagulation processes, and dust settling is treated by calculating the dust scaleheight which is  a function of grain size and local turbulence parameter \citep[for more details see][]{Gorti15} and for which we use a viscous parameter $\alpha =  5\times10^{-3} $. The dust temperature is computed for each grain (size and composition) at every $(r,z)$. Since the dust and gas are collisionally coupled with exchange of momentum and thermal energy, the dust and gas disk structures are determined together and a solution  is iteratively obtained  to be consistent with the constraints set by Equation~\ref{eq:vhse}. 

The density and temperature structure obtained at every $(r,z)$ of both gas and dust is used to solve for gas-grain chemistry.  
In this model, the ice surface and ice mantle are treated as two separate phases in interaction, and grain surface chemistry includes diffusion of the chemical species on the surface, two-body reactions, photoreactions, and thermal and nonthermal desorption mechanisms (see RG19 and references therein for more details). The gas-phase chemical network has been extended to account for HD chemistry and shielding \citep{Bergin13} and single and multiple isotope reactions of $^{13}$C and $^{18}$O. We include reactions listed in \citet{Roueff15} for $^{13}$C and \citet{Loison19} for $^{18}$O. For all other reaction rates we duplicate the main isotopologue network and assume similar rate constants and statistical branching ratios. Isotopic species are restricted to molecules with a maximum of three carbons and two oxygens. Selective photo-dissociation of CO isotopologues is considered using the self-shielding parameters from \citet{Visser09}. Finally, we assume similar binding energies for all ices formed from C and O isotopes. In total, the network consists of $\sim 800$ species and $\sim 10^4$ reactions.

Emission line fluxes are computed using an escape probability method which is the same as that used for the cooling calculations in the thermal balance computations and assuming face-on disks (zero inclination). Non-LTE radiative transfer includes collisional and radiative processes, and includes radiation from the dust background \citep[e.g.][]{Hollenbach79,Tielens85}, which is iteratively computed with the disk physical and chemical structure.   Molecular data are from the Leiden Atomic and Molecular Database (LAMDA). We use the escape probability method for the analysis of line fluxes as this allows us to compute the continuum background and treat spatial variations in dust opacities (relevant for HD emission); moreover, because this method is also used to compute cooling in the disk structure modeling we can directly compare the emission flux at a given location $(r,z)$ to the chemistry, heating and cooling processes. 

Using the above models, we conducted a limited parameter survey where the disk mass is varied in increments of 0.5 dex in the range $3 \times 10^{-4}-10^{-1}$M$_\odot$ (i.e. $0.3-100$ M$_J$) for three different dust/gas mass ratios $\epsilon=10^{-1},10^{-2}$ and $10^{-3}$. All these models assume an outer disk radius of 300au. For the canonical value of $\epsilon=10^{-2}$, we run an additional series where the outer disk radius is set to 100au, resulting in higher surface densities for a given mass when compared to the 300au series. In total, we run 24 different models.  In all cases, the stellar parameters are kept constant with $M_*=1.0M_{\odot}, R_*=2.0$ R$_\odot$ and UV and X-ray spectra similar to TW Hya (see RG19). We also assume that there is an ambient UV field with $G_0=1$ in Habing units. Cosmic ray ionization rates are depth dependent and calculated using the low energy fits derived by \citet{Padovani18}.  Table~\ref{tab:modelpar} lists the various assumed model parameters. As emission lines originate from the surface layers, where photoprocesses drive the chemistry on short timescales to cause a rapid reset, we assume all species are initially in atoms and ions at the start of the chemical simulation for computational efficiency. Chemistry in the disk may possibly retain some signature of the parent cloud core chemistry (see also RG19), but this is restricted to the inner midplane regions of the most massive disks and these regions are not emissive for our targeted lines, further justifying our assumption of the initial state.

\begin{table*}
    \centering
    \begin{tabular}{l c l c}
    \hline
    \hline
    Parameter & Value & Parameter & Value\\
    \hline
    Stellar Mass & 1.0M$_{\odot}$ & Stellar radius & 2.0 R$_\odot$ \\
    Stellar UV  & $10^{31}$ erg s$^{-1}$ & Stellar X-rays & $10^{30}$ erg s$^{-1}$  \\
    Dust silicate mass fraction & 0.83 & Dust carbon mass fraction & 0.17 \\
    Surface density function & $\Sigma(r)=\Sigma_0 (r_c/r) e^{-r/rc} $ & $r_c$ & 100 au \\
    Inner disk radius & 1 au & Outer disk radius & 100 and 300 au \\
    \hline
    Element 	& Abundance & Element & Abundance \\
    \hline
    He      & 0.1       &   N		& 6.2(-5) \\
    H		& 1.0 	    &   Si      & 1.7(-6)  \\
    D       & 1.6(-5)   &   Mg 		& 1.1(-6) \\
    $^{12}$C & 1.4(-4)  &   S       & 1.0(-6) \\
    $^{13}$C & 2.0(-6)  &   Fe		& 1.7(-7)   \\
    $^{16}$O & 3.2(-4)  &   PAH     & 1.00(-9) \\
    $^{18}$O & 5.74(-7) \\
    \hline
    \end{tabular}
    \caption{List of the assumed physical parameters for the disk models and gas-phase elemental abundances used in the chemical modeling. Gas disk masses and the dust/gas ratio are varied as described in the text.}
    \label{tab:modelpar}
\end{table*}
\section{HD and CO isotopologue line emission from models}
\label{sec:model_results}

Panels in the top row of Fig.~\ref{fig:coo} shows the physical structure of one of the model disks with a mass $10^{-2}$ \ms\  and a dust/gas ratio $\epsilon=10^{-2}$.  HD and CO isotopologue emission typically originates in regions where the $A_V$ to the star is between $\sim 1-10$. Densities in this region are $\sim 10^{7-9}$ cm$^{-3}$ and gas and dust temperatures are quite similar for the model disk shown in Fig.~\ref{fig:coo}. The average dust cross-section per H is reduced by a factor of nearly 100 in the $1<A_v<10$ region compared to the ISM value of $\sim 5\times 10^{-21}$ cm$^{-2}$ due to the larger dust grain sizes in disks and due to settling. The disk structure varies with disk gas and dust masses as discussed in RG19 in greater detail.

In the disk regions responsible for production of the line emission discussed here, gas is strongly coupled to the dust via collisions with smaller, hotter dust grains heating the gas and larger, colder grains cooling it. Dust collisions are typically a net coolant for the more massive disks with additional gas heating provided by FUV grain photoelectric heating, and gas is warmer than the dust. For the lower mass disks, dust collisions are typically a net heating source  with additional cooling provided mainly by CO and [OI]63$\mu$m emission and in this case gas is slightly cooler than the dust.  For details on the included heating, cooling processes and how they affect density/temperature distributions, please see \citet{Gorti11} and RG19.

\subsection{HD and CO isotopologue chemistry}
\label{sec:chemistry}

\begin{figure*}
    \centering
    \includegraphics[width=18cm]{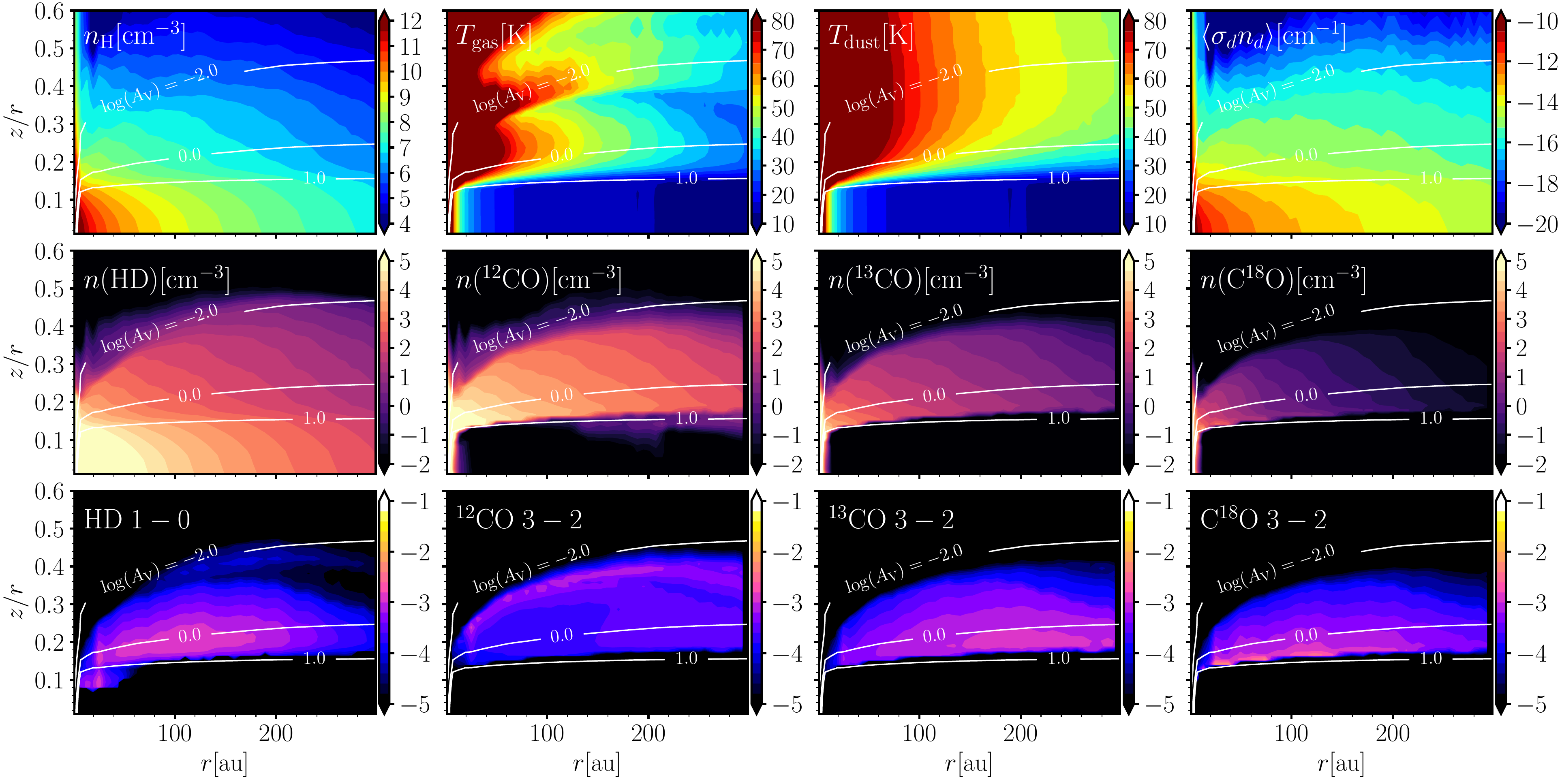}
    \caption{\label{fig:coo} Computed physical structure, density maps and line luminosities of a selection of molecules for a model disk with \eps$ = 10^{-2}$ and $M_\text{gas} = 10^{-2} M_\odot$. The top panel shows the gas density ($n_\text{H}$), gas temperature ($T_\text{g}$), the area-weighted mean dust temperature ($T_\text{d}$) and the dust cross-sectional area and the dust density averaged over the grain size distribution ($\langle \sigma_\text{d} n_\text{d} \rangle $) as a function of the disk radius $r$ and the normalized height $z/r$. Solid line contours show the visual extinction ($A_V$) to the star. The middle and lower panels show the computed density maps of HD, \coc\ and \coo\ and the predicted line emission map 
    normalised by the total emission of the line (J$=1-0$ for HD, J$=3-2$ for CO isotopologues).}
\end{figure*}

Figure~\ref{fig:coo} also shows the computed density and emission maps of HD and the three CO isotopologues considered here. Although the H/HD transition occurs at lower $z$ than the H/H$_2$ transition, HD chemistry is similar to that of H$_2$.  HD forms on grains and is photodissociated by UV photons; other reactions involving D or D$^+$ exchanges are not found to be important in setting the abundance of HD in the disk. Gas phase HD abundance is also not affected by grain surface chemistry due to the low binding energy of the HD molecule on grains and HD is therefore abundant even in the midplane of the disk. 

\begin{figure}
    \centering
    \includegraphics[width=0.47\textwidth]{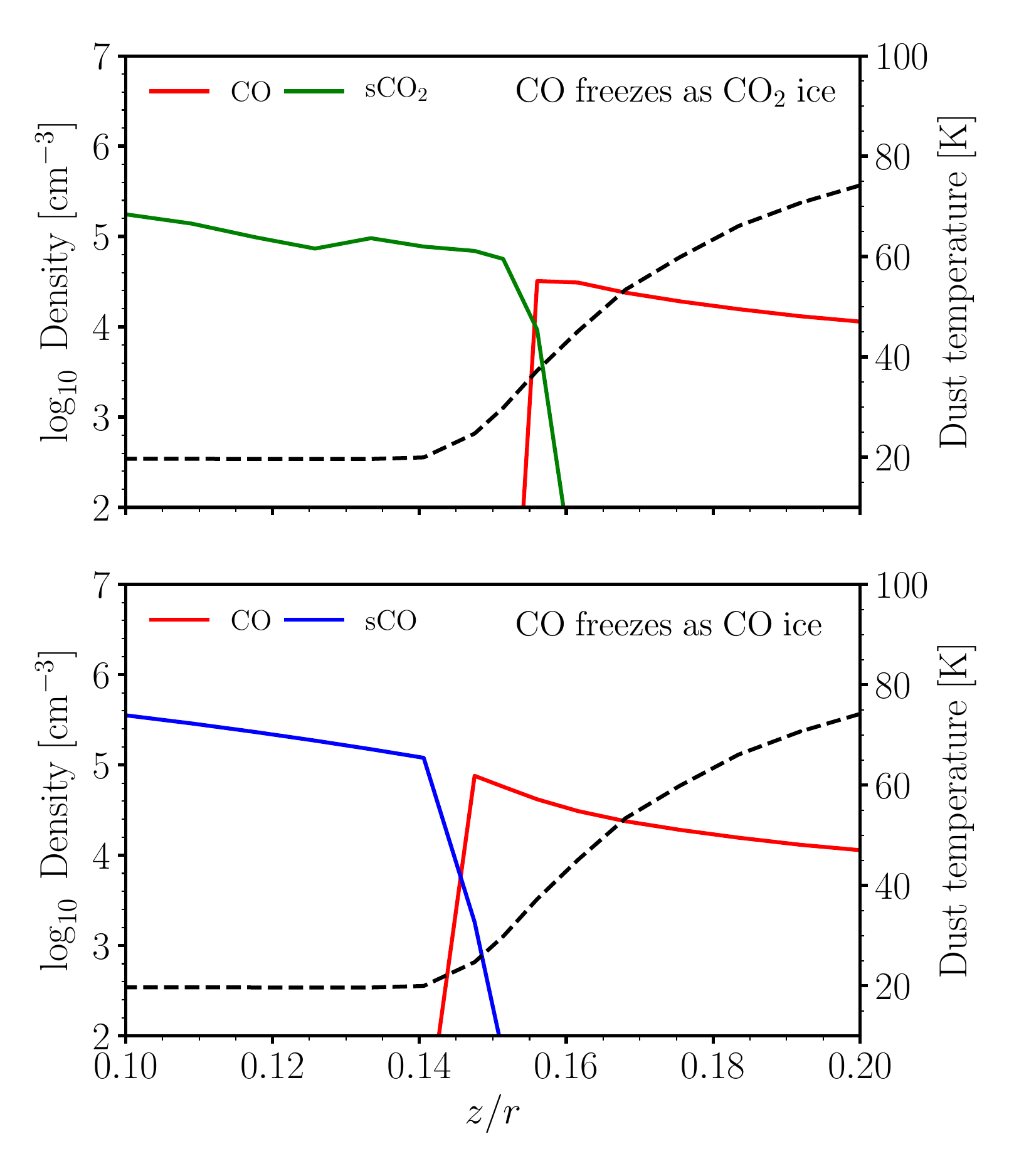}
    \caption{\label{fig:noconv} Computed vertical gas-phase density of CO and ice density of CO and CO$_2$ (denoted as sX) as a function of $z/r$ and at $r=50$ au of the disk shown in Fig. \ref{fig:coo}. The top panel shows results when CO conversion to CO$_2$ ice was considered and the bottom panel when this conversion was neglected. Black dashed lines show the dust temperature. When CO conversion into CO$_2$ ice is included, the location of the CO snowline is at higher $z$ because of the higher binding energy of CO$_2$ relative to CO. For this disk and at this radius, this effect results in all CO being converted to CO$_2$ ice even in the midplane due to cosmic ray ionization \citep[see][]{Ruaud19}, and  a  $\sim 25\%$ decrease of the vertical gas CO column density as compared to the case where this conversion is neglected.} 
\end{figure}

The vertical column densities of gas phase CO and its isotopologues are determined by the location of the CO photodissociation front at $A_v \sim 10^{-2}$ and the height where water ice begins to form at $A_v \gtrsim 2-3$. Gas-phase CO is present in between these two layers (see Fig.~\ref{fig:coo}). In RG19, we show that photoprocessing of ice, near the surface corresponding to the vertical water snowline, induces efficient conversion of CO into CO$_2$ ice and impacts the location of the vertical CO snowline. In this process, photodissociation of water ice near its snowline leads to the production of OH which rapidly reacts with CO on grains  (before it can desorb) to form the more bound CO$_2$ ice. The location of the CO snowline is thus determined not by the CO desorption/condensation balance but by desorption of CO$_2$ that the CO ice converts to. The vertical CO snowline is therefore shifted higher up from the midplane because of the higher binding energy of CO$_2$ relative to CO, and the CO snowline more or less coincides with the location of the vertical water snowline which is itself determined by photodesorption \citep[see][for similar conclusions]{Furuya14,Aikawa15,Molyarova17,Bosman18,Trapman2021}. This is illustrated in Fig. \ref{fig:noconv} which shows a comparison of the vertical chemical structure computed when this effect is taken into account and when it is neglected. The gas density at the CO snowline location depends on the dust cross section per H nucleus which can be quite low in disks due to both grain growth and settling.  Along the radial direction, and in the inner disk midplane, UV photons are produced by cosmic rays and the low UV field results in a CO to CO$_2$ conversion on longer ($\sim$ Myr) timescales. As a result, the CO snowline slowly moves inward until most of the CO has been converted to CO$_2$ ice in a few Myr\footnote{As will be discussed in \S 5, the location of the snowline as a function of time can be obtained by solving the reduced network given in Appendix \ref{sec:co_snowline_loc}} (for more details, see RG19). 

The vertical extent of gas phase CO and its isotopologues is determined not only by their condensation front but also by where they are photodissociated in the surface layers. In the disk atmosphere (i.e. at $z/r\gtrsim0.4$ for the model shown in Fig.~\ref{fig:coo}), all CO isotopologues are subject to photodissociation by stellar and interstellar FUV radiation. Self-shielding prevents photodissociation of CO and its isotopologues and they begin to rise in abundance when $z/r\lesssim0.4$.  \coc\ chemistry is not particularly sensitive to selective photodissociation and is closely coupled to \co\ chemistry through the exchange reaction

\begin{equation}
    ^{13}\mathrm{C}^+ + \mathrm{CO} \leftrightarrow \mathrm{C}^+ + {^{13}\mathrm{CO}} + \Delta E = 35 K
\end{equation}

which enhances the abundance of \coc\ compared to \co\ (relative to that expected from selective dissociation effects alone). 
However, isotope selective photodissociation has a significant impact on the distribution of \coo\ which is not shielded by \co\ or \coc\ and is abundant only in a thin layer located just above the vertical CO snowline (at $z/r\sim0.15$ for the model shown in Fig.~\ref{fig:coo}) where it forms by reaction of $^{18}$O with CH and CH$_2$.

\subsection{HD and CO isotopologue line emission}
\label{sec:line_emission}
The lower panels of Figure \ref{fig:coo} show contributions, as a function of spatial location, to the corresponding emission of HD $J=1-0$ and the $J=3-2$ line of the CO isotopologues. For CO isotopologues we focus on the emission of the $J=3-2$ transition which is usually brighter in disks. 

HD traces the main constituent H$_2$ and is abundant throughout the disk, including the dense midplane regions; however, emission from the $1-0$ line arises mainly from the warmer upper layers of the disk, with a small region of emission in the inner disk where the gas is sufficiently warm. In both these regions, one at the inner disk midplane (at $z/r\lesssim 0.1$ and $r\lesssim 20$ au in Fig.~\ref{fig:coo}) and the other above the disk midplane (at $z/r\gtrsim 0.15$ and $r\gtrsim 20$ au), the temperature of the gas ranges between 30 K and 80 K. As has been noted by previous authors \citep{Bergin13,Trapman17,Kama20}, the relatively higher energy level of the HD $1-0$ transition ($E_u \sim 128$K) makes HD emission quite sensitive to the gas temperature at the cold temperatures typical of disks because the fractional population in the $J=1$ level, and hence emission of the $J=1-0$ line, is proportional to $\exp(-128/T_\text{gas})$. HD is moreover suppressed by the strong continuum background in disks at $112\mu$m, which lowers the line flux that can be detected above the continuum. There is thus a relatively sharp drop in HD emission at the height where the dust temperature transitions from the optically thin value \citep[the so-called superheated layer, e.g.,][]{chiang97}. The HD $J=1-0$ line therefore is sensitive to the local temperature and density structure and originates mainly from the surface layers of the disk.

Emission of all the CO lines is also confined to the surface, and specifically to the spatial regions between the CO snowline and the photodissociation fronts where CO and its isotopologues exist in the gas phase. As is to be expected, \co\ is optically thick and the $J=3-2$ transition shown in Fig.~\ref{fig:coo} peaks in a narrow layer near the CO photodissociation front.  Depending on the disk mass and gas column density, \coc\ $J=3-2$ could be optically thin but for the model disk shown here the line is marginally optically thick. As discussed earlier, \coc\ is not significantly affected by selective isotope photodissociation and the emitting region is therefore more extended. The line luminosity is dominated by the regions closer to the snowline and with $r \gtrsim 150$ au for this model disk. Contrary to \coc,\  \coo\ emission mostly originates from regions with $r\lesssim$ 150 au, where the density is higher, and is optically thin at most radii even for disks more massive than the $10^{-2}$\ms\ disk shown here. The emission, like that of HD, is therefore weighted toward higher densities (lower $z$) and thus higher column densities, and regions located just above  the CO snowline dominate the emergent flux. We note that the gas temperature, which is within a few K of the dust temperature, is typically $\gtrsim 30$K above the CO snowline where CO is in the gas phase. An important consequence therefore is that while both HD and \coo\ luminosities are sensitive to the gas density, unlike HD, the lower energies of the \coo\ sub-mm transitions make  \coo\ luminosities relatively insensitive to the disk temperature structure.

\subsection{Disk mass parameter survey}
\label{sec:mass_survey}

\begin{figure*}
    \centering
    \includegraphics[width=18cm]{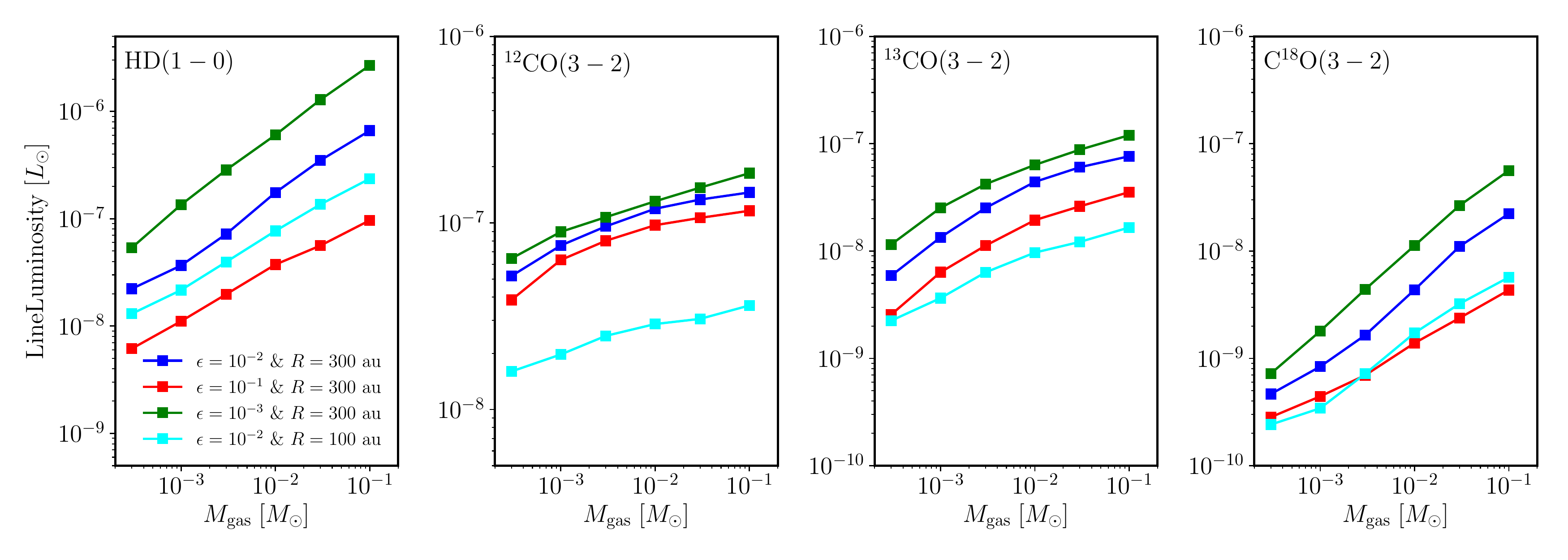}
    \caption{\label{fig:all_runs_emmission_vs_gas_mass} HD (1-0),\co\ 3-2, \coc\ (3-2), \coo (3-2) emission as function of gas mass for different dust/gas ratios ($\epsilon=10^{-2}$:blue, $\epsilon=10^{-3}$:green and $\epsilon=10^{-1}$:red). Cyan color data show 
    the $\epsilon = 10^{-2}$ model disks 100au in radial extent and are to be compared with the blue $\epsilon = 10^{-2}$ dataset with radii of 300au. The differences in emission for the more optically thin lines are mostly due to differences in columns in the emitting regions. For optically thick lines such as \co\ and \coc\ (which is marginally optically thick), differences are mainly due to line optical depth.}
\end{figure*}

Figure \ref{fig:all_runs_emmission_vs_gas_mass} shows the computed luminosity of HD ($1-0$), \co, \coc\ and \coo ($3-2$) lines for all the models of the parameter survey discussed in \S \ref{sec:model_desc} as a function of the gas disk mass. We consider three values of the dust/gas ratio; this is fixed as a function of radius (i.e., $\Sigma_\mathrm{dust}(r)/\Sigma_\mathrm{gas}(r) = \epsilon = 10^{-1}, 10^{-2}$ and $10^{-3}$) but varies with height due to dust settling. We also consider one set of models with $\epsilon=10^{-2}$ but with a smaller radial extent. All line luminosities exhibit an increasing trend with gas mass, with a slope that is shallow for the more optically thick \co\ and \coc\ lines and steeper for HD and \coo. \co\ emission is relatively insensitive to the dust/gas ratio, mainly because it originates high above the disk (see fig.~\ref{fig:coo}) at unit line optical depth where the gas temperature is determined by FUV and X-ray heating mechanisms and line cooling. Because the emission is optically thick, the CO luminosity depends on the disk spatial extent and decreases substantially for the series with smaller disk size. \coc\ line emission is marginally optically thick and in general exhibits trends intermediate to very optically thick \co\ emission and optically thin HD and \coo\ emission. Emission from HD and \coo\ is optically thin at most disk radii; the former depends on the column density of warm HD and the latter on the column density of CO above the vertical snowline respectively, and these in turn depend on the disk physical structure as we describe below.

\paragraph{Gas mass variations for a fixed $\epsilon$:} 
The optically thin emission does not scale linearly with gas disk mass (for a fixed $\epsilon$) as might be expected because the local gas density and resulting column densities also depends on the disk structure---denser gas is in general cooler which affects the emission.
This was previously noted by \citet{Trapman17} who also found a sublinear change in HD emission with gas mass. \coo\ emission, being less sensitive to temperature by virtue of its lower excitation temperature as previously discussed, exhibits a steeper trend with gas mass compared to HD, although still sub-linear (i.e. \coo\ 3-2 luminosity approximately scales with $\propto M_\text{gas}^{0.7-0.8}$ while HD 1-0 luminosity is $\propto M_\text{gas}^{0.6-0.7}$). The emitting regions for HD and \coo\ approximately lie near the $A_V=1$ layer in the disk, at a few scaleheights. While the overall density in the disk is proportionately lowered when the disk mass is decreased, a shift in the emitting regions to denser gas at lower $z$ compensates to produce the sublinear trend with disk mass. This shift is mainly due to decreased dust/gas coupling in less dense disks; this causes greater settling of dust to move the $A_V=1$ layer and affects the difference in dust and gas temperatures.

\paragraph{Line luminosities with $\epsilon$ variations:}
The disk surface ($A_V\sim 1$ layer) and flaring are sensitive to the disk structure, which is again affected by the dust/gas coupling which is determined by $\epsilon$. The gas temperature is nearly the same as the (optically thin) dust temperature at the surface in the dense, thermally well coupled, HD and \coo\ emitting regions and does not vary much for the different $\epsilon$ models.For a given disk mass, however, high values of $\epsilon$ move the emitting layer to higher $z$ and lower gas densities affecting the optically thin emission lines of HD and \coo. For the $M_\text{gas}=10^{-1}$\ms\ disk, both the HD and \coo\ luminosities increase by $\sim30$ when $\epsilon$ changes from $10^{-1}$ to $10^{-3}$. At lower gas masses, this trend with $\epsilon$ continues for the HD line. However, \coo\ line luminosity slope as a function of gas mass is shallower for dustier disks (red curve in Fig.~\ref{fig:all_runs_emmission_vs_gas_mass}), because in these disks the increased dust (relative to gas) contributes more toward shielding \coo\ from FUV photodissociation, thereby increasing the abundance of \coo\ and producing more line emission.

\begin{figure*}
    \centering
    \includegraphics[width=0.9\textwidth]{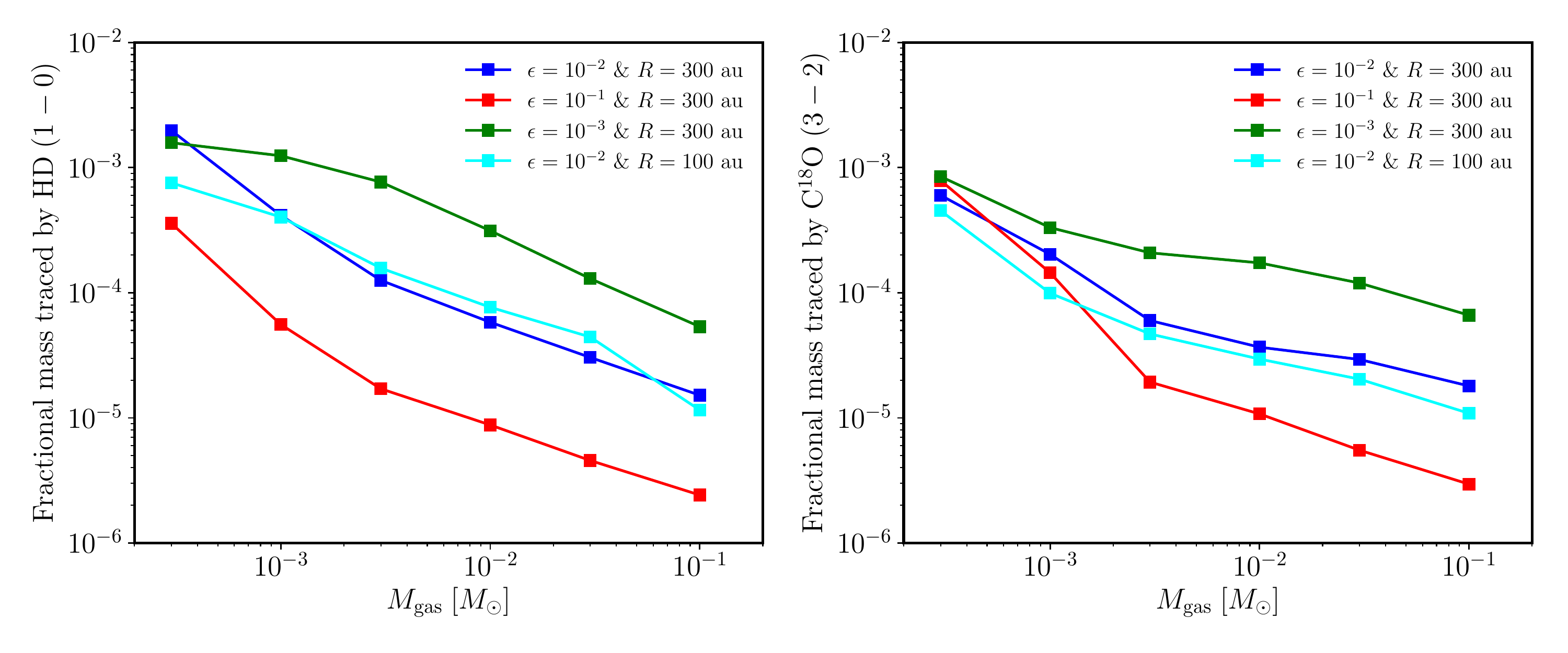}
    \caption{\label{fig:mass_traced_c18o_hd} Gas mass fraction traced by HD and C$^{18}$O as a function of the total gas mass of each disk, calculated by weighting the mass in a given volume element by the fractional contribution to the line luminosity from that element. For both lines, the mass traced is typically less than $\sim 10^{-3}$ or $0.1$\% of the total disk mass for most of the cases.}
\end{figure*}

\paragraph{Fraction of gas mass traced by HD and \coo \ line emission}
We next compute the fraction of the gas mass that is traced by the HD (1-0) and \coo\ (3-2) emission, i.e., we sum the gas mass in each cell at $(r,z)$ by weighting it by the fractional luminosity contribution from that cell. These results are shown in Fig.~\ref{fig:mass_traced_c18o_hd}. The most striking result apparent from these figures is that HD and \coo\ both trace a small fraction ranging between $\sim 10^{-6}-10^{-3}$ of the total gas mass for all the various disk models considered here. This is mainly due to the fact that the dense midplane is not emissive; most of the HD and almost all the \coo\ emission comes from the disk surface. If the optically thin dust disk surface ($\sim A_V=1$) is at some height $z_t=\chi h$, where $h$ is the scaleheight corresponding to the temperature at the midplane, then the fraction of mass contained above this height ($z=z_t$ to $z=\infty$) is equal to $\Sigma_s (z>z_t)/\Sigma = 1- \mathrm{erf}(\chi/\sqrt{2})$ which is $\sim 5\times10^{-7} - 3\times10^{-3}$ for values of $\chi = 3-5 $. This mass fraction again depends on the disk structure (which determines $\chi$ and $h$) and the trends observed in Fig.~\ref{fig:mass_traced_c18o_hd} can be traced back to those in Fig.~\ref{fig:all_runs_emmission_vs_gas_mass}, because lower disk masses in general result in lower values of $\chi$.

Our results therefore indicate that \coo\ is just as effective in tracing the disk mass as HD, and in fact could be a better tracer. HD and \coo\ emission both depend on the gas density, but \coo\ is, in addition, relatively insensitive to the temperature which is $\sim 30-80$K  in the emitting regions, i.e., HD 1-0 emission is proportional to $\exp(-128/T_\text{gas})$ while \coo\ 3-2 lies at $E_u\sim 32$K, and is proportional to $\exp(-32/T_\text{gas})$. Conversion of HD flux to total gas mass, on the other hand, requires that the temperature  structure of the gas be sufficiently well known. Further, multiple rotational \coo\ lines are readily accessible by ALMA, NOEMA, SMA and other sub-mm facilities, breaking some model degeneracies, whereas the HD (1-0) line at 112$\mu$m is currently inaccessible. We note here that the results presented above do not invoke any CO depletion factors and assume standard values (as in the ISM) for the elemental C and O abundances. We next compare our results with observations, our discrepancies with several previous works, and possible reasons.

\section{Comparison with ALMA surveys and previous chemical modeling analyses}

\subsection{Agreement with ALMA CO surveys}
\label{sec:alma_comp}

\begin{figure*} 
    \centering
    \begin{tabular}{ll}
    \includegraphics[width=0.45\textwidth]{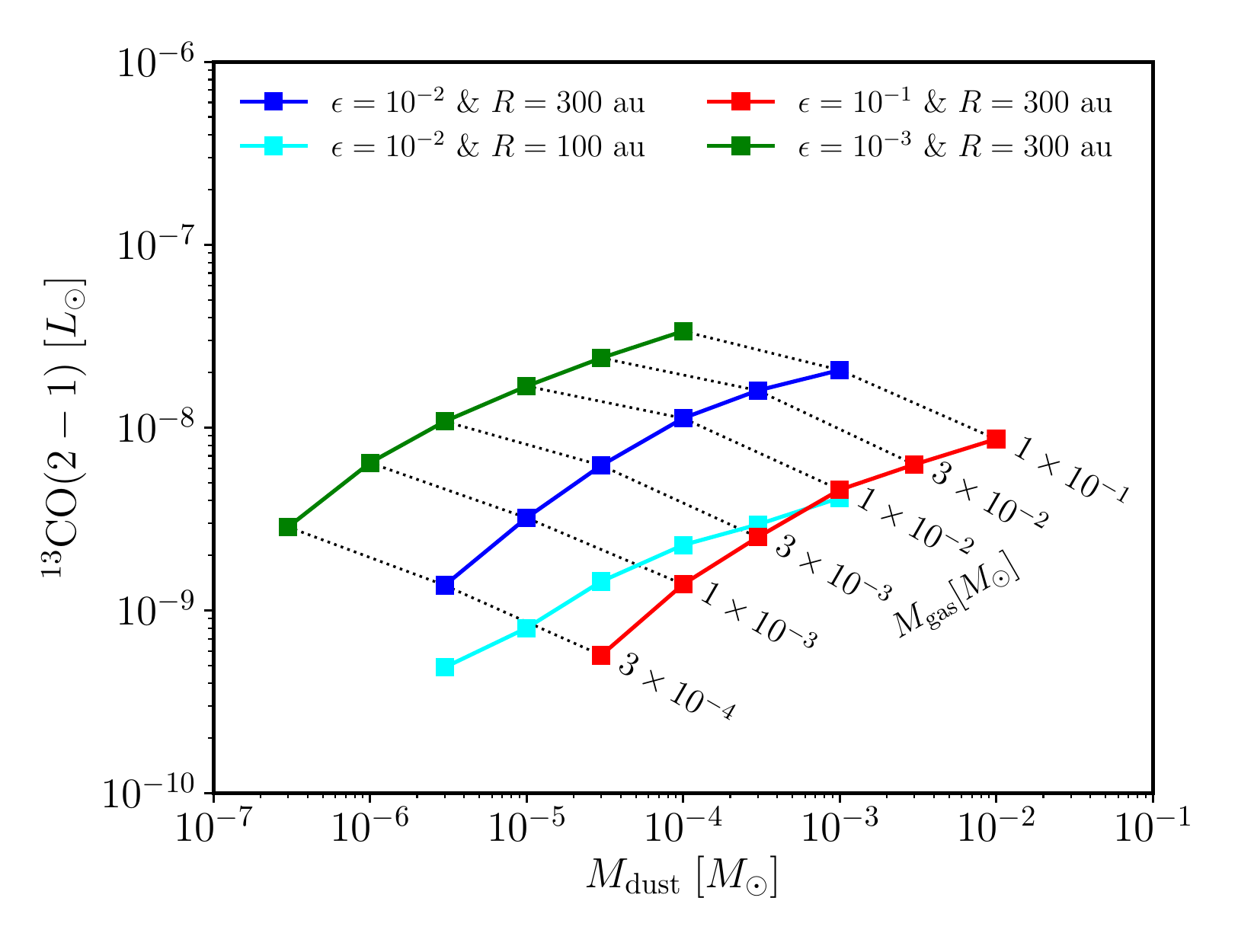} &
    \includegraphics[width=0.45\textwidth]{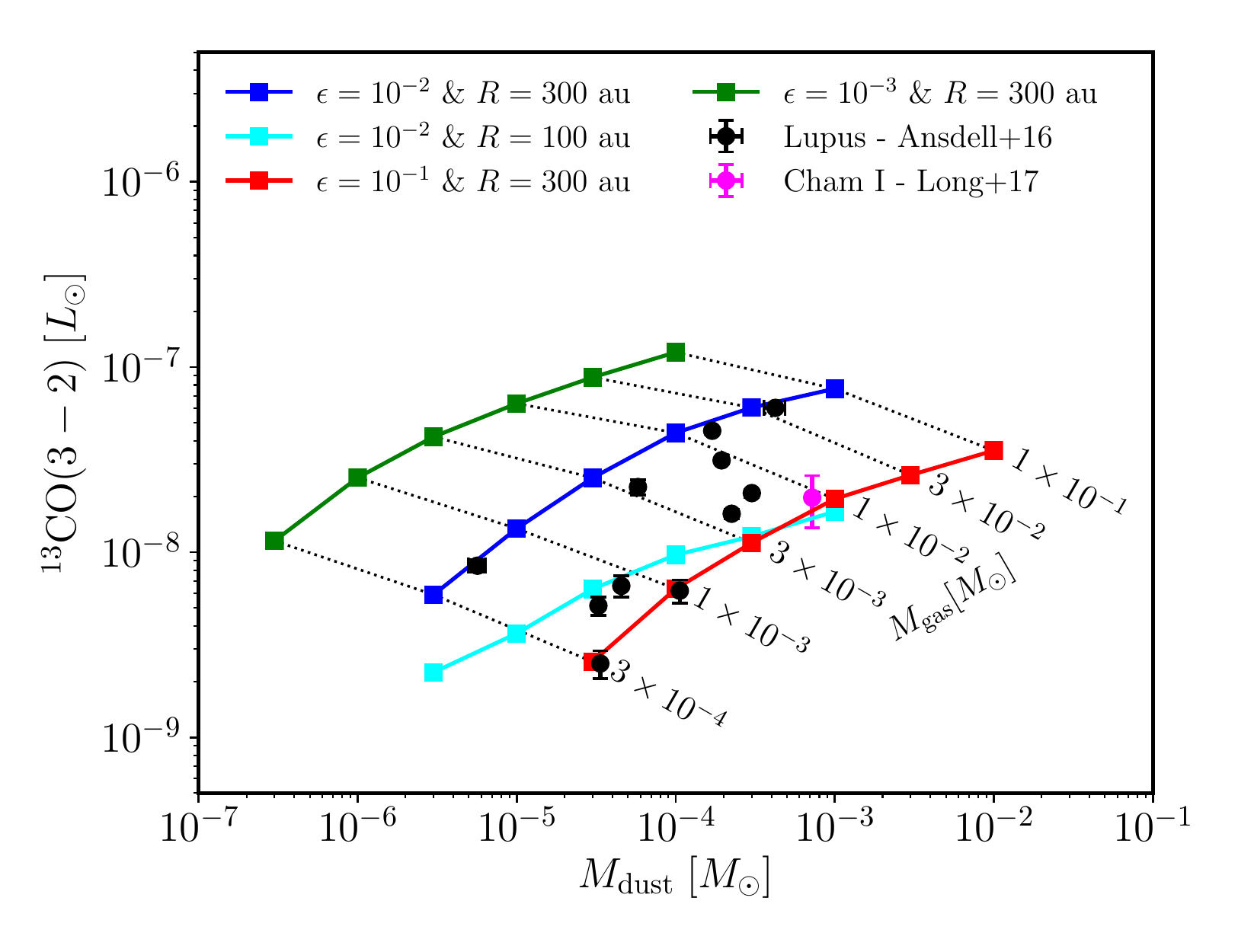}
    \end{tabular}
    \caption{\label{fig:mdust_13co}$^{13}$CO (2-1) and (3-2) emission as function of dust mass for different dust/gas ratios, assuming zero inclination for the disks. Black points show \coc\ (3-2) results obtained from the Lupus survey \citep{Ansdell16}. The point in magenta shows a result obtained from the Cham I survey \citep{Long17}. The cyan model points are for
    a smaller 100 AU disk with $\epsilon=10^{-2}$.}  
\end{figure*}

\begin{figure*} 
    \centering
    \begin{tabular}{ll}
    \includegraphics[width=0.45\textwidth]{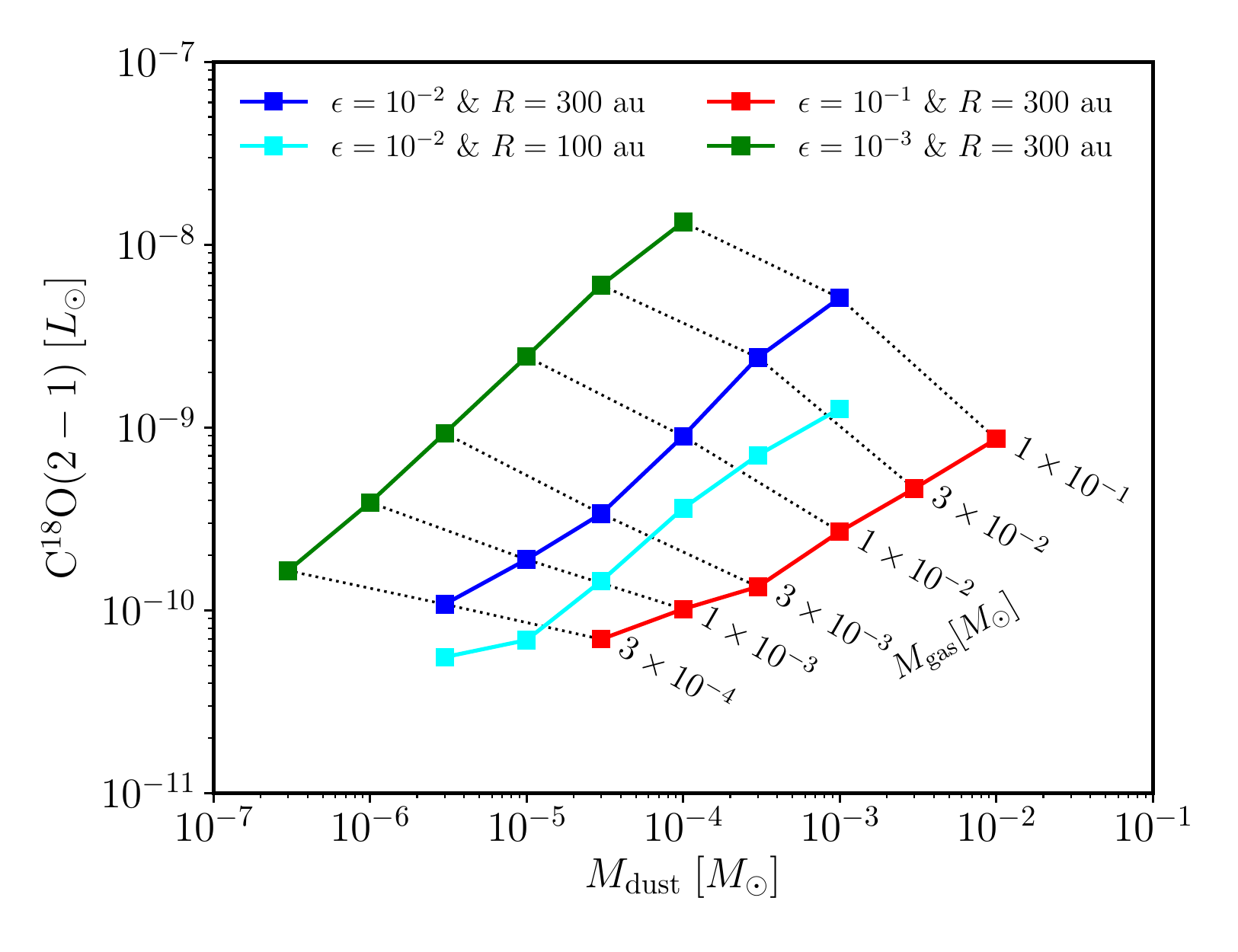} &
    \includegraphics[width=0.45\textwidth]{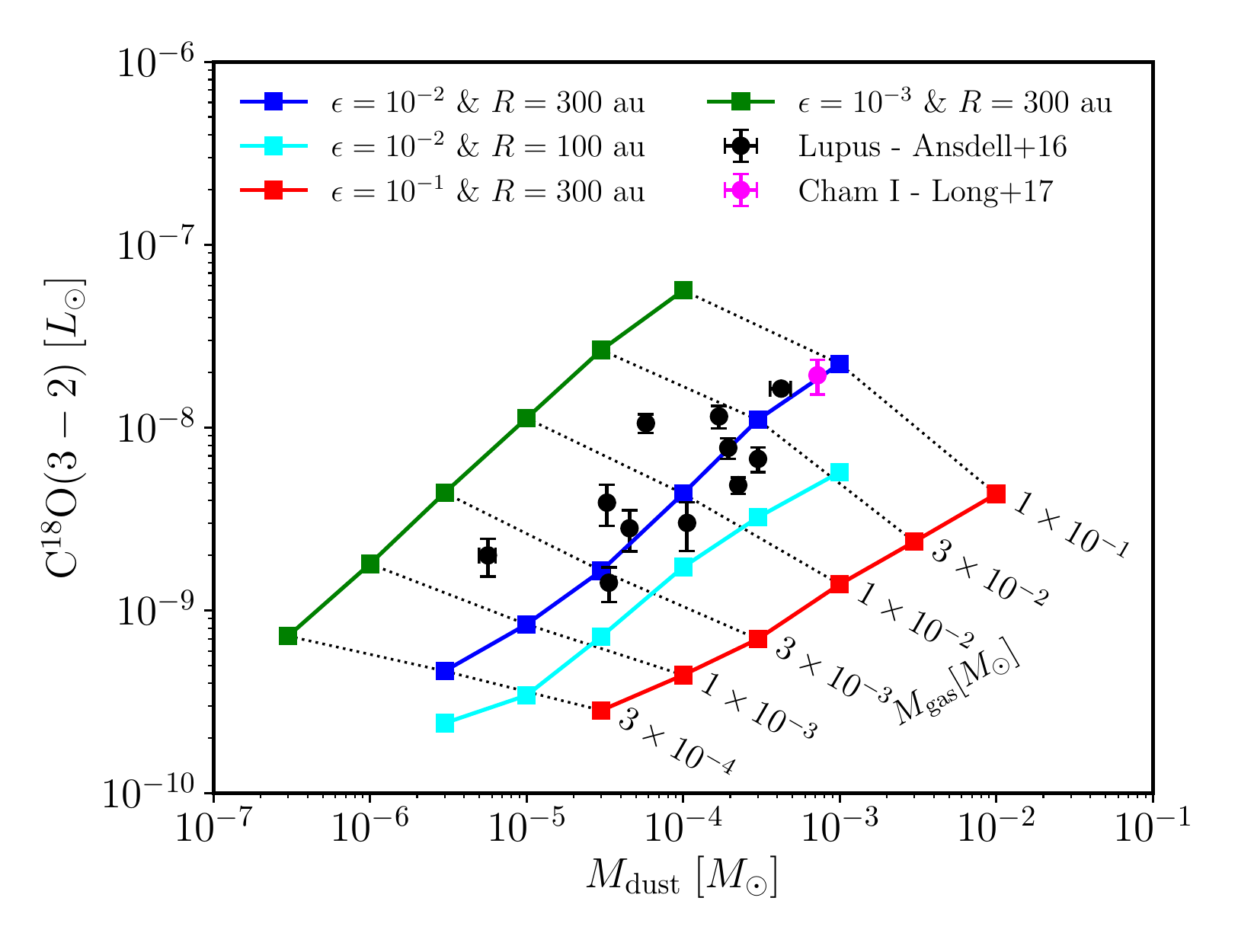}
    \end{tabular}
    \caption{\label{fig:mdust_c18o}C$^{18}$O (2-1) and (3-2) emission as function of dust mass for different dust/gas ratios. Black points show \coo\ (3-2) results obtained from the Lupus survey \citep{Ansdell16}. The point in magenta shows a result obtained from the Cham I survey \citep{Long17}. The cyan model points are for a smaller 100 AU disk with $\epsilon=10^{-2}$.}
\end{figure*}

In Fig. \ref{fig:mdust_13co} and \ref{fig:mdust_c18o} we compare our computed CO luminosities with luminosities derived from surveys in Lupus \citep{Ansdell16} and Chamaeleon I \citep{Long17}, which are plotted as a function of the inferred mm dust masses. We focus on disks for which the potential gas mass tracer \coo\ has been detected, that is to say, 11 disks for the Lupus survey \citep{Ansdell16} and only one disk for the Chamaeleon I survey \citep{Long17}. Note that \co\ luminosities from these surveys are not available in the literature. Of all the model series, the disk models that reproduce observations the best are those with a normal interstellar dust/gas ratio and ISM like  elemental abundances. For this set of models, the model predicted line luminosities of \coc\ and \coo\ can explain the ALMA data to within a factor of a few. However, note that while dust masses for the models are input parameters and hence known, we have simply used quoted dust disk masses to locate observational data in the figures and that these include inherent uncertainties in the continuum flux to dust mass conversion \citep[by up to a factor of $\sim 5$, see e.g.][]{Ballering19,Tychoniec2020,Ribas20,Macias21}. Comparison of \coc\ and \coo\ luminosities show that the models slightly overpredict \coc\, or underpredict \coo\, by a factor of $\sim 2$. Observed line ratios give \coc/\coo$\sim 2-4$ while modeled ratios vary between $3-10$ for all the model series. 

Without access to \co\ observations in these disks, which could have helped in distinguishing between these two possibilities, we first explore the possibility of \coc\ overprediction. Our investigations show that it is extremely difficult to decrease \coc\ emission by a factor of $\sim 2$ by modifying chemistry  without also affecting the emission of \co\ and \coo. 
In the models, \coc\ chemistry is completely dominated by the exchange reaction with \co\ \citep[see also e.g.][]{Miotello14}. Varying input parameters like stellar UV flux, X-ray luminosity, metal abundances and PAH initial abundances, or disk surface density distributions, did not affect this result. Part of the \coc\ overprediction compared to ALMA data can be explained by the fact that our model flux computations assume face-on, zero inclination disks, while there are a range of inclinations likely present in the observed survey samples. Whereas inclination will not affect the optically thin \coo\ line, the \coc\ line is marginally optically thick (see Fig.~\ref{fig:all_runs_emmission_vs_gas_mass}) and inclined disks are expected to have lower observed luminosities. However, disk inclination is likely to suppress \coc\ by less than a factor of 2 on average, and even if this is included, our models will still overestimate \coc\ by a small factor.

Another possibility for decreasing \coc\ emission while keeping \coo\ emission roughly constant would be to truncate the disk. As shown seen in Fig. \ref{fig:coo}, \coc\ emission originates from the outer disk (i.e. $r\gtrsim 150$ au) while \coo\ emission originates in regions with $r\lesssim 150$ au. However, as seen in seen in Fig. \ref{fig:mdust_13co} and \ref{fig:mdust_c18o} for disks with $\epsilon=10^{-2}$, decreasing the disk radii from $R=300$ au and $R=100$ au decreases \coc\ emission by a factor $\sim 3-5$ but it also decreases \coo\ emission by a factor of $\sim 2-3$. Thus a more compact disk can only partly explain the factor of $\sim 2$ overprediction in the \coc\ emission. Other possibilities include currently missing chemical pathways able to compete with the exchange reaction with \co\ and thus lower \coc\ abundance in the disk (see also Section \ref{sec:twhya}). We note that recent work by \citet{Trapman2021} also find that \coc\ is over-produced by a factor of $\sim 2$ by their models.

Investigations on the possibility of \coo\ underprediction show that increasing \coo\ emission by a factor $\sim 2$ could be achieved by small variations in the chemistry. Only a small fraction of the available $^{18}$O is in the form of \coo\ in the models and  photochemistry dominates its destruction. As a result, a small increase in the formation rate of \coo\ can help in increasing \coo\ column density and resulting emission. As discussed in \S 3, \coo\ is predominantly formed by reaction of  $^{18}$O with CH and CH$_2$ and this rate has not explicitly been measured. In our model we assume that the rates of reaction with $^{18}$O are similar to that with $^{16}$O, for which rate constants are available only at room temperature or above \citep{Loison14}. A test for the $\epsilon=10^{-2}$ model series shows that increasing these rates by a factor of 2 is enough to enhance \coo\ emission by a factor of $\sim 1.5$. The conversion rate of CO into CO$_2$ is also uncertain; in this case, decreasing the conversion of CO into CO$_2$ ice could increase the abundance of gas phase CO close to the vertical CO snowline and result in an increase in the \coo\ column density and associated emission. A test on the $\epsilon=10^{-2}$ models shows that completely neglecting the conversion of CO ice into CO$_2$ ice enhances \coo\ emission by a factor of $\sim 1.5-2$. Finally, because photochemistry dominates the destruction of \coo, uncertainties in the \coo\ shielding factors could also affect the computed \coo\ emission.

To summarize, unlike the conclusions from associated modeling in previous studies \citep[e.g.][]{Ansdell16,Miotello17}, our results can recover the observed \coc\ and \coo\ line fluxes without invoking increased dust/gas ratios or CO depletion factors and are only discrepant by a small factor of $\sim 2$ from the observations. As figure \ref{fig:mdust_c18o} shows, \coo\ emission can be explained with disks having ISM like elemental abundances  (see Table~\ref{tab:modelpar}) for C and O, and with ISM like dust/gas ratios, $\epsilon = 10^{-2}$.

\subsection{Discrepancies with earlier models }
\label{sec:model_comp}

Our results substantially differ from those obtained from other similar thermochemical studies, and here we investigate possible reasons for these differences. Some of the earliest modeling \citep[e.g.,][]{Favre13} calculated \coo\ fluxes higher than observed and reported CO being underabundant by more than two orders of magnitude. Later studies showed that isotope selective photodissociation affects the CO isotopologue fluxes, and that CO underabundances (or alternately, gas depletion) could be only about an order of magnitude \citep[][]{Miotello16,Miotello17}.  However, our results do not indicate any order-of-magnitude discrepancies. 

While it is difficult to directly assess the key differences without formal bechmarking due to the complexity of the codes, we attempt to identify some differences below which apply to many currently used thermochemical codes including DALI \citep[][]{Bruderer12}, ProDiMO \citep[][]{Woitke2016} and RAC2D \citep[][]{Du15}. Our analysis revealed the following main differences: 
\begin{itemize}
\item[-] Most models (including DALI, ProDiMO and RAC2D) use a fixed isothermal Gaussian profile for the density structure of the disk in computing  hydrostatic pressure equilibrium and do not update this to be consistent with the chemistry and  their computed gas temperatures. 
\item[-] The dust disk structure is not coupled to the vertical gas density or pressure profile, instead key quantities like the scaleheight and flaring angle are parameterized independently.
\item[-] In many cases, a simplified grain surface chemistry is used where only accretion, desorption and hydrogenation reactions are taken into account. CO is assumed to freeze on grains when dust is at $\sim 20$K, with no additional chemistry. Some models further do not treat selective photodissociation of CO isotopologues.
\end{itemize}

We next investigate how the above assumptions affect the calculated \coo\ model fluxes. 
\coo\ emission is mostly optically thin, and therefore, the luminosity in a line is determined by the population of the higher level of the transition. For the $J=3-2$ line, the upper level population follows $N_{J=3} = N_\mathrm{C^{18}O} \exp(-32 \text{K}/T_\mathrm{gas})/Z(T_\mathrm{gas}) \propto N_\mathrm{C^{18}O} \exp(-32/T_\mathrm{gas})/T_\mathrm{gas}$; where $N_\mathrm{C^{18}O}$ is the total number of \coo\ molecules and $Z(T)$ is the partition function. For typical emitting regions $T_\mathrm{gas}\sim 20 - 100$K, and the level population and hence line luminosity, peaks near the excitation temperature ($T \sim 32$K) and weakly declines with gas temperature. The emission is, therefore, relatively insensitive to gas temperature. The column density of \coo\ molecules is consequently the only likely factor  responsible for the differences in \coo\ emission in our models compared other studies.

As discussed in \S \ref{sec:line_emission}, the \coo\ column density is set by the CO snowline (at $z_\mathrm{ice}$) and the \coo\ photodissociation layer (at $z_\mathrm{PDR}$) and is equal to 
$\int_{z_\mathrm{ice}}^{z_\mathrm{PDR}} n_\mathrm{C^{18}O}\ dz$.
\citet{Miotello14} studied the effects of isotope selective photodissociation and determined that this effect reduces the column density of \coo\ (by lowering $z_\mathrm{PDR}$) and resulting emission by almost an order of magnitude.
The dominant contribution to the \coo\ columns however arises from the lower heights due to the rapid decline in density with $z$. For models that include isotope selective photodissociation, the location of the CO snowline and the density here are therefore the main discrepant factors.

The significance of the CO snowline is evident; the assumption that CO freezes at dust temperature $\sim$ 20K  vs.  35K due to conversion to CO$_2$ ice moves the snowline deeper into the disk midplane where densities are high and will result in a clear over-estimate of the \coo\ column density at a given radius (see Fig.~\ref{fig:noconv}) and over-estimate the \coo\ luminosity. We find that this chemical conversion effect is one key difference between our models and earlier results.  Recent DALI models  by \citet{Trapman2021} that include CO conversion also find a factor of $\sim 3-5$ difference in the \coo\ and \coc\ fluxes from earlier DALI models by \citet{Miotello17}.

\begin{figure*}
    \centering
    \includegraphics[width=0.8\textwidth]{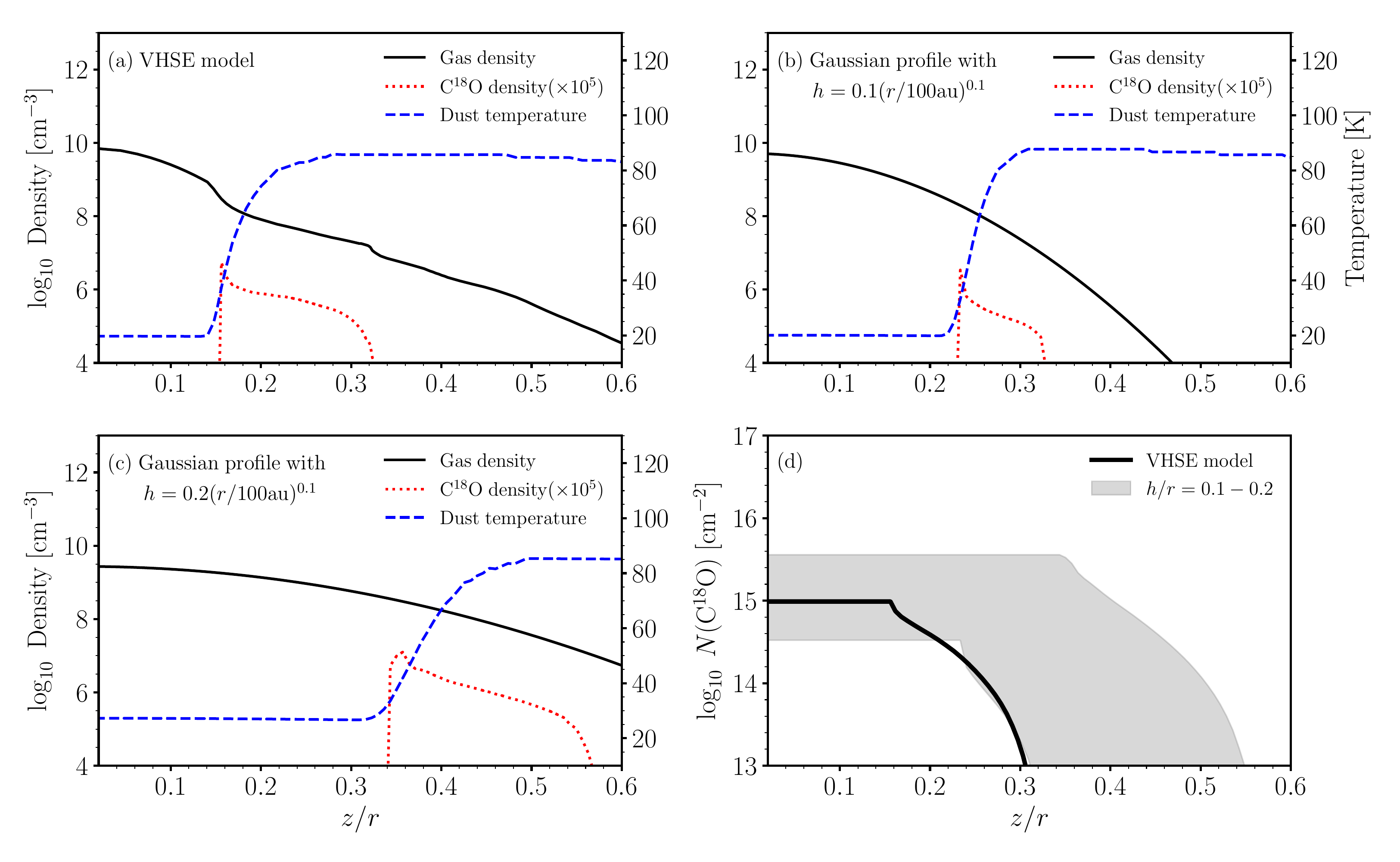}
    \caption{\label{fig:comp_miotello} Computed vertical structure as a function of $z/r$ and at $r=50$au of a disk with $\epsilon = 10^{-2}$ and $\Sigma_\text{gas}(r=50\mathrm{au}) \sim 2.0$ g.cm$^{-2}$ obtained with our full thermo-chemical disk model (a), and a parametric expression which assumes that the vertical density structure follows a Gaussian profile with $h/r=0.1$ (b) and $h/r=0.2$ (c).  The black solid lines are for the gas density, the blue dashed lines for the dust temperature and the red dotted lines for the computed C$^{18}$O density. The last panel (d) shows the range of vertical \coo\ column density as a function of height computed for these different models. The effect of the conversion of CO into CO$_2$ ice at the surface of the grains is taken into account for all these models.}
\end{figure*}

More fundamentally, the location of the snowline is in itself a function of the disk structure.  Figure ~\ref{fig:comp_miotello}
shows the comparisons at $r=50$\ au between our vertical hydrostatic equilibrium model (VSHE, top left panel) and 
models where the density profile is a fixed gaussian with
two choices of parametrization.  The density profile is assumed to be determined by a power law form for the scale height $h$ and two values typically adopted in the literature ($h/r \sim 0.1$ and $h/r \sim 0.2$) are used. The density distribution for the two gaussian profiles (top right, $h/r=0.1$; bottom left, $h/r=0.2$) are evidently different from the pressure equilibrium profile (top left panel of Fig.~\ref{fig:comp_miotello}). Further, the dust temperature transitions at different heights in the disk depending on the adopted value of the scaleheight. These differences result in marked changes in $z_\mathrm{PDR}$ (due to density) and $z_\mathrm{ice}$ (due to dust temperature profile) and therefore very different \coo\ densities when compared with our self-consistent models. The bottom right panel of Fig.~\ref{fig:comp_miotello} shows the corresponding range in column density as a function of height which is seen to be quite sensitive to the adopted parametrization of the density profile. We thus find that calculating the disk density structure with reasonable accuracy, in tandem with the dust radiative transfer and in a self-consistent manner is important for calculating the \coo\  line luminosities.

The HD emission lines in our model are also systematically lower by a factor $\sim 2$ as compared to the emission calculated by \citet{Trapman17} using the DALI code, and this again may be directly related to differences in disk physical structure. HD emission also arises from above the $A_V=1$ layer where our densities are lower as described above, but is in addition sensitive to the column density of warm gas (unlike \coo\ emission which is only sensitive to column density). Since the bulk of the disk gas is cold, differences in the temperature (e.g. mean HD emitting temperature of 45 K vs. 35K) could account for the factor of two difference in emission between the models. 
HD emission is further sensitive to other factors like the dust/gas ratio, which affect the disk physical structure and the detectability of the line above the continuum. The disk structure in our model is strongly coupled to disk properties (surface density of gas and dust are the only varied parameters), but many quantities including the dust scale height are treated as independent variable parameters in \citet{Trapman17} making direct comparisons with our work difficult. We note however that in a test model, \citet{Trapman17} also find a factor of 2 decrease in the computed HD (1-0) emission when the density and temperature structure are iterated a few times by solving the equation of hydrostatic equilibrium (see their Appendix C, note they do not run these models to convergence). We therefore conclude that the main difference can be attributed to their isothermal density structure and our density/temperature coupling through the hydrostatic equilibrium condition.

\subsection{The TW Hya disk and HD vs. CO discrepancies}
\label{sec:twhya}
\begin{figure*}
    \centering
    \includegraphics[width=1.0\textwidth]{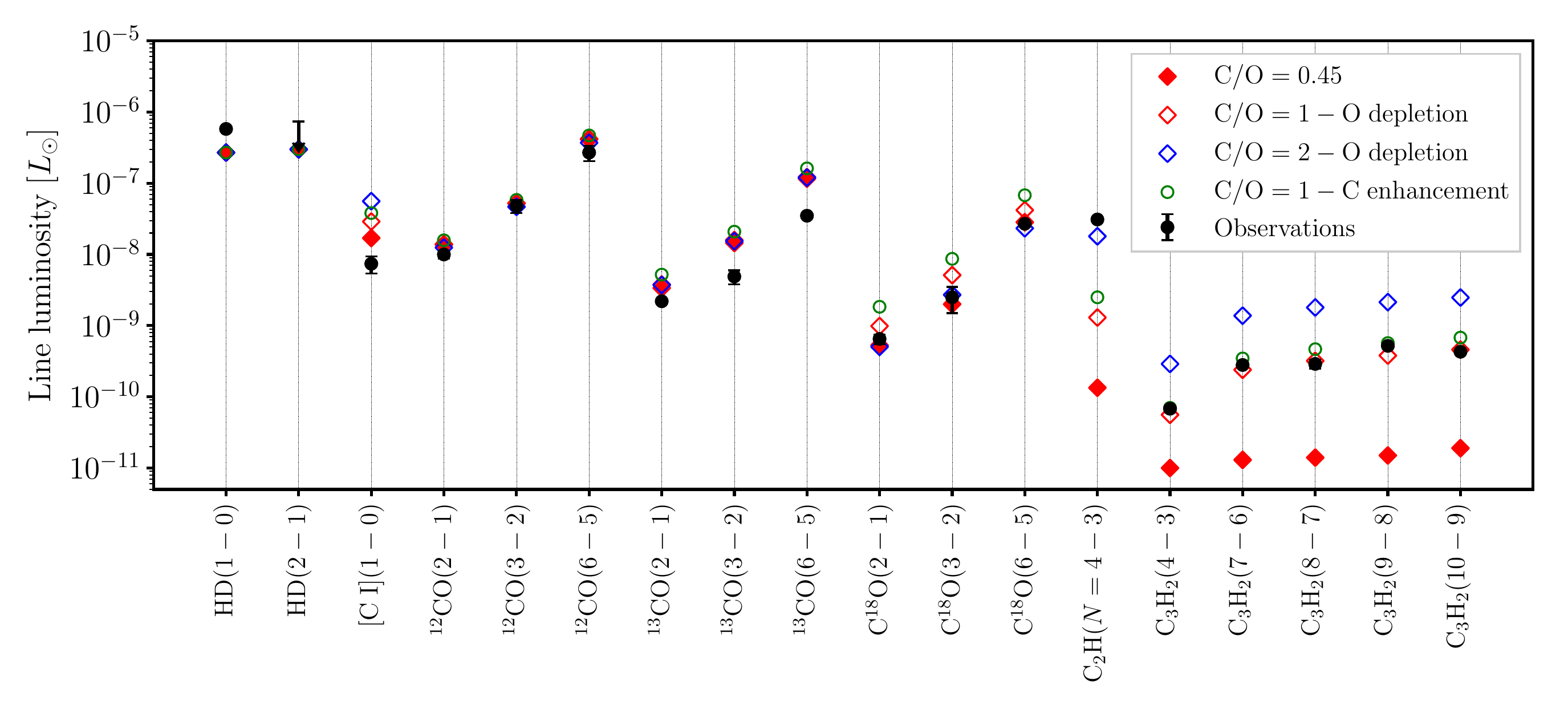}
    \caption{\label{fig:twhya_comp} Computed line luminosities for TW Hya and comparison with observations of HD \citep{Bergin13}, CO \citep{Schwarz16,Calahan21}, C$_2$H \citep{Bergin16}, C$_3$H$_2$ \citep{Cleeves21}. The reference model (C/O=0.45) assumes ISM-like elemental abundances, i.e. C/H = $1.4\times 10^{-4}$ and O/H = $3.2\times 10^{-4}$.}
\end{figure*}

Most arguments for CO depletion and underabundances of both C and O come from the copious observational data available for the disk around the nearby star, TW Hya \citep[e.g.,][]{Bergin16}. This is the only disk for which HD has been detected with sufficient S/N, and for which there is therefore a chemistry-independent assessment of gas mass \citep[][]{Bergin13}. In this section we compare our results with observations for the TW Hya disk using the structure determined in \citet{Gorti11} using the thermo-chemical model of \citet{Gorti04,Gorti08} which solves for vertical hydrostatic equilibrium. This model was based on the dust model  by \citet{Calvet02} and reproduces the spectral energy distribution (SED) and can match  the integrated emission of several surface emission lines including H$_2$ S(1) and S(2), HCO$^+$ (4-3), CO rotational lines, CO rovibrational lines, OH line emission, and [Ne II], [Ne III], [O I] fine structure emission and  optical forbidden lines, as well as the HD flux observed with {\it Herschel} \citep[][]{Bergin13,Mcclure2016} to within a factor of $\sim2-3$.
In \citet{Gorti11},  no depletion factors for C or O were assumed, and normal ISM-like elemental abundances along with the standard value of $\epsilon=10^{-2}$ were used. Using this disk physical structure (densities and temperatures), here we compute gas grain chemistry as in RG19 and line luminosities for \co, \coc, \coo\ and a few other species (Fig. \ref{fig:twhya_comp}) that were not previously considered.

With the notable exception of hydrocarbons, to be discussed later, the the new chemical model reproduces observed CO isotopologue lines to within a factor of a few despite using standard values for $\epsilon$ and the elemental C and O abundances. 
\co\  and \coo\ fluxes are very well matched (see Fig.~\ref{fig:twhya_comp}), but as in the Lupus survey comparison there is systematic overestimation of the $^{13}$CO (3-2) and (6-5) emission by a factor of $\sim 2-3$ (note TW Hya is nearly face-on, so inclination does not explain this discrepancy, see discussion in \S \ref{sec:chemistry}).

HD $1-0$ emission is under-estimated by a factor of $\sim$ 2, as in earlier work \citep{Bergin13} that used the same \citet{Gorti11} disk structure, and there is thus an inconsistency of a factor of $\sim 2$ with respect to the \coo\ emission. Previous authors \citep[e.g.][]{Favre13,Du15,Kama16,Trapman17,Bergin18} have, however, invoked large depletion factors ranging from $10-100$ of carbon and oxygen to reconcile HD and CO observations in TW Hya. For the various reasons discussed in \S \ref{sec:model_comp}, abandoning parametrization of the disk structure and adopting vertical hydrostatic equilibrium derived density/temperature distributions may alleviate the need for such severe depletion of C and O. Our results nevertheless indicate that CO isotopologue emission and HD emission could be discrepant, but by a small factor of $\sim 2-3$ at most. To some extent, the lower HD could be mitigated assuming a slightly enhanced elemental abundance of D  \citep[we use $X(\mathrm{D})=1.6\times 10^{-5}$  compared to $X(\mathrm{D})\sim 2\times 10^{-5}$ used in e.g][]{Trapman17,Calahan21}. Considering dust settling \citep[which was not treated in][]{Gorti11} may also slightly decrease the dust background and increase the line flux over the continuum. We note that we did not recompute the TW Hya disk model with settling as this would involve reconciling the spectral energy distribution with the gas disk structure again and involves considerable effort. This will be the subject of a forthcoming study. Another possibility is that the disk mass is higher. However, the system is an old ($\sim$ 8 Myr) transition disk with evidence for planet formation and possibly embedded planets \citep[e.g.][]{Teague19}. Disk masses higher than the $0.06$\ms\ indicated here would moreover imply this evolved disk is gravitationally unstable with a mass comparable to the central star. We therefore consider significantly higher gas disk masses to be unlikely, but note that a mass $\sim 2$ times higher than in our model has been inferred by \citet{Powell19} with a different approach using resolved multiwavelength continuum data.

While our models with ISM elemental abundances of C and O (and C/O $\sim 0.45$) can successfully explain most line emission data \citep[also see][]{Gorti11}, including HD, calculated emission from C$_2$H and C$_3$H$_2$ is far lower than observed. 
A change in the C/O ratio to a value $>1$ is a straightforward explanation as this leaves free carbon (not bound in CO) available for hydrocarbon production as expected from chemical considerations; C/O$>1$ has thus been invoked in order to explain abundances of hydrocarbons inferred from observations \citep{Bergin16,Kama16}.
As seen in Fig.~\ref{fig:twhya_comp}, increasing the C/O ratio via oxygen depletion (i.e. C/H=O/H=$1.4\times10^{-4}$) or carbon enhancement (i.e. C/H=O/H=$3.2\times10^{-4}$) both give nearly similar results for all molecules and only marginally affect the CO isotopologue line emission. Assuming C/H=O/H=$1.4\times10^{-4}$ slightly increases \coo\ emission because lowering O decreases free oxygen available to form water ice and affects the conversion of CO to CO$_2$ ice on grain surfaces as it also lowers the abundance of the OH radical. This shifts the vertical CO snowline to lower $z$ and results in an higher column density of CO and thus higher \coo\ emission (see \S \ref{sec:chemistry}). A further decrease in oxygen to a C/O ratio of 2, decreases the abundance of \coo\ throughout the disk due to limited availability of oxygen and slightly decreases \coo\ emission. Enhancing carbon to affect the C/O ratio results in slightly higher \coo\ luminosities due to an overall increase in CO isotopologue abundance.  \co\ and \coc\ lines are optically thick in this disk and  are not affected by changes in the C/O ratio.

Hydrocarbon emission, on the other hand, is drastically affected by changes in the C/O ratio, with a factor of $\gtrsim 10$ increase in line luminosities due to the increased availability of carbon. This is due to the fact that the hydrocarbon abundances needed to produce the observed emission are only a small fraction of the elemental carbon available (e.g., $X(\text{C}_2\text{H}) \sim 10^{-8} \sim 10^{-4} X(\text{C})$). While C$_3$H$_2$ emission is well explained with C/O=1 (which is independent of the assumption made on either enhancing carbon or depleting oxygen), C$_2$H emission is still under-produced by both the C/O $\sim$ 1 models. Increasing C/O to 2 results in an improvement in matching  C$_2$H emission, but significantly over-produces C$_3$H$_2$ emission. We can thus constrain any possible change in the C/O ratio in the TW Hya disk to $\sim1-2$, which causes only in a small change in the CO isotopologue emission lines.

The main conclusion that can be drawn from the models is that lowering the elemental abundances of C and O by large factors is not required to explain line emission from the TW Hya disk, and hydrocarbon emission favors a possibly small change in the C/O ratio; this could moreover be restricted to the surface of the disk where emission lines originate. Furthermore, unlike that of carbon chains, \coc\ emission is only marginally affected by a change in the C/O ratio. 


We note, however, that spatially resolved line observations are available for the TW Hya disk, and many of the models invoking elemental depletion \citet{Zhang17, Zhang19, Calahan21} have attempted to recover the radial profile of CO isotopologue emission with parametric fitting. Data obtained from the recent ALMA MAPS program \citep{Oberg21b}, and in particular constraints obtained in favorably inclined disks also allow inferences on the vertical origin of the emission \citep[e.g.][]{Zhang21,Calahanb,Schwarz21}. For a more definitive conclusion on the  extent to which the C/O ratio changes in disks we need to self-consistently model the structure of individual disks in greater detail to simultaneously reproduce all the infrared line emission and all spatially resolved sub-millimeter line emission; this will be subject of future work.

\section{Simplified models for estimating gas disk masses using \coo\ line fluxes}
\label{sec:simple_model}

We have thus far argued that \coo\ lines are reliable tracers of the disk gas mass; the analysis presented in \S 4 demonstrated that \coo\ emission traces a similar total disk mass fraction as HD, is less sensitive to uncertainties in disk structure and temperature, and that consistent physico-chemical modeling indicates discrepancies are limited to a factor $\sim 2$ when compared to HD for the TW Hya disk. However, our models are inordinately complex and estimating disk masses from observations by detailed modeling of disks (as for TW Hya) is impractical. From our model results, we have identified two main modeling components that need to be considered in order to infer spatial density distributions and estimate gas masses to within a factor of a few from \coo\ emission lines (and \coc\ lines):
\begin{itemize}
    \item[-] A physical model that provides a reasonable estimate of the vertical density structure at the $A_V\sim1-3$ layer near the water condensation front. 
    \item[-] A minimal chemical network including the chemistry of CO isotopologues, selective photodissociation, freeze-out and conversion of CO into CO$_2$ ice at the surface of grains.
\end{itemize}
\subsection{Simplified vertical density structure of the disk}
Determining the disk physical structure requires solutions to the equations of vertical pressure balance coupled to the equations of radiative transfer. Pressure is determined by gas temperature and density, and the gas temperature is further coupled to the chemistry and dust properties, making consistent disk solutions a formidable computational effort.  Therefore, the problem becomes much more tractable by decoupling the density structure calculation from the chemistry. Models often simplify the procedure by assuming a parametric Gaussian (gas assumed isothermal) vertical density profile that is kept fixed, and the gas temperature and chemistry are iterated to convergence. Very often, when modeling individual disks, the procedure adopted is justified by finding a dust temperature that can reproduce observed continuum spectral energy distributions (SEDs). However, the dust temperature is not isothermal with height and transitions from a cooler midplane to a warmer surface near the $A_V=1$ layer \citep[see several older works, for e.g.,][]{Calvet91, chiang97, dalessio98,  Dullemond2001}. 
As discussed earlier (also see RG19), most emission lines originate above $A_V=1$ with excitation determined by the temperature at the surface, while the midplane temperature characterizes the regions where most of the gas mass resides and sets the gas vertical density structure. Therefore, characterizing the density profile with a single gaussian characteristic of the surface emission will lead to an underestimate of the density in the cooler midplane regions (see Fig. \ref{fig:comp_miotello}). We therefore advocate against parametrizations based on the notion of a single scaleheight.

For modeling \coo\ isotopologue emission, in particular, the density structure is more critical than the temperature structure (see \S \ref{sec:alma_comp}) and we find that a dust disk model can reasonably approximate the full solution \citep[see e.g.,][ for a similar approach]{Mcclure2016, Molyarova2017} and provide a useful alternative approach. We advocate a density structure based on the dust temperature profile that is consistent with pressure equilibrium. We use the publicly available Monte Carlo radiative transfer code RADMC-3D \citep{Dullemond12} and follow a very simple procedure to illustrate this method. For an input surface density distribution, we set an initial guess density distribution with height and calculate the temperature with RADMC-3D. In our simple setup, we use a size averaged dust opacity which yields a single effective (radiative) dust temperature and equate this to the gas temperature. The density distribution is re-calculated according to the vertical pressure gradient and re-normalized to the surface density (see Eq.~\ref{eq:vhse}). The new density distribution is used again as input to RADMC-3D to get a new temperature distribution and this process is repeated to convergence. We note that dust settling is not taken into account in this procedure although we find that the resulting change to the dust structure does impact the C$^{18}$O by a small factor.

\subsection{Simplified CO chemistry}
The disk physical structure (i.e. density and temperature as function of $(r,z)$) obtained as described above can then be used to compute the CO chemistry. The minimal chemical network for estimating disk gas masses from CO isotopologue observations needs to include H$_2$ and CO isotopologue gas phase chemistry, self-shielding, selective photodissociation, CO freeze-out and subsequent conversion of CO into CO$_2$ ice at the surface of grains. For H$_2$ and CO chemistry, we find that the reduced network proposed by \citet{Gong17} for hydrogen and carbon chemistry gives good results when compared with our full network. It includes 18 chemical species coupled through $\sim 50$ reactions, including photoreactions as well as self- and mutual-shielding for H$_2$ \citep{Draine96} and CO \citep{Visser09}. This network then needs to be extended to take into account the chemistry of $^{13}$C and $^{18}$O and in particular reactions listed in \citet{Roueff15} for $^{13}$C and \citet{Loison19} for $^{18}$O. For all other reactions, the chemical network can simply be duplicated assuming similar rate constants and statistical branching ratios. Isotope selective photodissociation can reduce \coo\ emission by up to a factor of $\sim 10$ \citep{Miotello14} and therefore needs to be included; shielding functions are provided in \citet{Visser09}. Finally,  we find that $\sim 10$ reactions are sufficient for CO grain surface chemistry. The set of reactions is given in Appendix~\ref{sec:co_snowline_loc} and takes into account CO freeze-out and its subsequent conversion into CO$_2$ ice as well as water ice formation and ice photochemistry, which are both important to determine the location of the vertical CO snowline. 

\subsection{\coo\ emission from simplified models}

\begin{figure*} 
    \centering
    \begin{tabular}{ll}
    \includegraphics[width=0.45\textwidth]{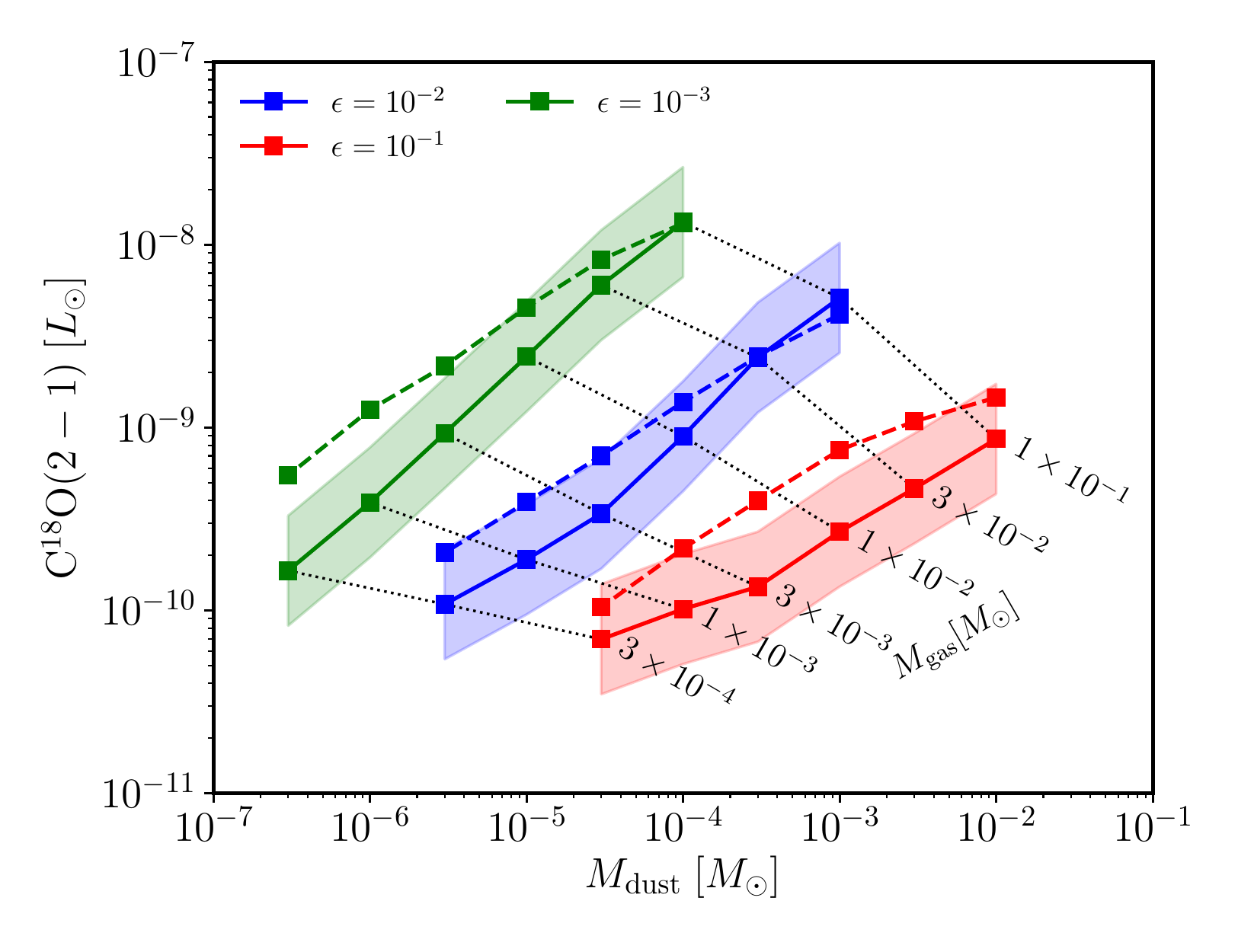} &
    \includegraphics[width=0.45\textwidth]{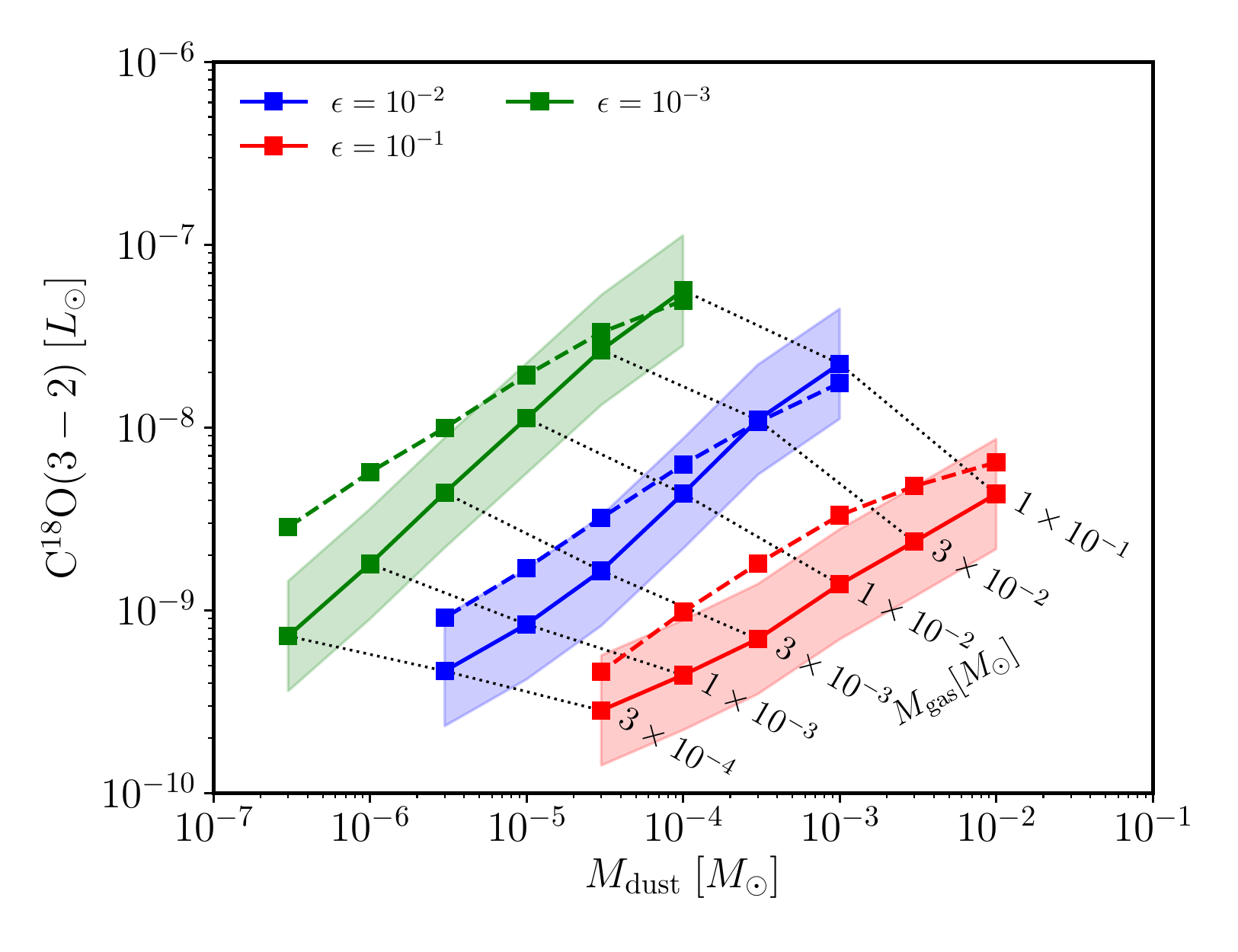}
    \end{tabular}
    \caption{Comparison between C$^{18}$O (2-1) and (3-2) emission computed with the full thermo-chemical model (solid lines) and the simple RADMC-3D-based dust disk model (dashed lines), for different dust/gas ratios. The filled region denotes a factor of 2 deviation from the emission obtained with the detailed model. For most disks, results obtained from the simple model lie close or within this region.}
    \label{fig:c18o_dust_model}
\end{figure*}

In Fig. \ref{fig:c18o_dust_model} we show a comparison of the \coo\ emission computed using the full thermo-chemical model and the simple  model described above. For most of the disks explored here, the use of a simple dust model leads to $\lesssim 2-3$ difference in the computed \coo\ (2-1) and (3-2) emission compared to our full thermo-chemical model, and provides a computationally efficient and reasonably accurate procedure to retrieve gas disk masses from observational data. The reason that this procedure works reasonably well is because \coo\ emission mainly arises from regions where  dust and gas are collisionally coupled and where gas temperature deviates only slightly from that of the dust.
The RADMC-3D-derived model is found to over-estimate the \coo\ emission slightly, due to differences in density caused by differences in the true gas temperature from the dust mean radiative value and consequently the column density of \coo\ in the disk. 

We emphasize that while the above procedure may be a good approximation to compute emission lines from layers close to the water snowline, it fails at greater heights where gas and dust decouple and their temperatures deviate, and is not appropriate for modeling \co, \coc, as well as near and mid-infrared emission lines. The disk structure also depends on the dust properties (grain size distribution, opacity, settling) and uncertainties in these will consequently affect derived gas disk masses as well. Nevertheless, we expect the above method to be a good approximation for gas masses.

\section{Conclusions} 
Using self-consistent thermo-chemical gas and dust disk models coupled with detailed chemical models including dust settling and gas-grain chemistry, we reassess the potential of using CO isotopologue emission for tracing gas masses of protoplanetary disks. A parameter survey was conducted by varying gas disk masses, dust/gas mass ratio and disk outer radius to study the effects on the computed CO isotopologue and HD emission. Our main conclusion is that, using appropriate disk models, \coo\ line emission can be used to infer gas disk masses to accuracies comparable to what is achieved by disk dust mass estimations. \coo\ lines are readily accessible with ALMA, can further be spatially resolved to retrieve the surface density distribution, and multiple rotational transitions can be used to constrain the gas temperature. In addition to its accessibility, \coo\ is moreover not sensitive to the typically unknown gas temperature structure of the disk, and traces a similar disk mass fraction compared to HD, making it a better indicator of disk mass.
Our results can be summarized as follows:
\begin{enumerate}
    \item The disk physical structure has a significant impact on the computed line emission of CO isotopologues and HD. In particular, in order to determine gas masses using \coo\ emission lines it is necessary to make a reasonable estimate of the vertical density structure just above the CO freeze-out layer where most of the emission originates. We argue that parametrized density profiles are inadequate and pressure equilibrium needs to be considered in disk structure models.
    \item We find that HD and \coo\ both trace only a very small fraction of the total gas mass ranging from $\sim 10^{-6} - 10^{-3} $ for all the various disk models considered here. This is because both HD and \coo\ emission originate from regions above the disk midplane where most of the mass resides.
    \item Based on the above finding, we conclude that \coo\ is a overall a good tracer of the gas disk mass for typical disk conditions.  HD and \coo\ emission both depend on the gas density, but contrary to HD emission which shows an exponential dependence to the temperature of the gas, \coo\ emission is relatively insensitive to the temperature in the emitting layer. 
    \item An important finding of our study is that we can explain CO isotopologue line emission from surveys using ISM-like elemental abundances, i.e. C/H = $1.4\times 10^{-4}$ and O/H = $3.2\times 10^{-4}$, and a normal interstellar dust/gas ratio, i.e. $\epsilon=10^{-2}$. We reach the same conclusion for the TW Hya disk, where modeled HD and \coo\ emission are only discrepant by a factor of $\sim2$. This result substantially differs from those obtained by other studies where strong depletion factors of C and O are usually invoked. We find that the assumption of an isothermal profile for the density can lead to large deviations in the density at \coo\ emitting heights and therefore also the computed column densities and hence line emission.
    \item In agreement with previous studies, we also find that explaining the observation of species like small hydrocarbons in TW Hya seems to require a factor of $\sim 2$ enhancement in elemental carbon or a factor of $\sim 2$ reduction in elemental oxygen as compared to the ISM, with their ratios ranging between C/O$\sim 1-2$. We also note that hydrocarbon emission  does not provide any information on a potential global depletion of carbon and oxygen but only on their relative abundance (i.e. C/O) ratio. 
    \item Given that detailed models of disks as those presented here are complex, we provide a simplified approach to computing \coo\ line fluxes and determine gas disk masses using a dust-derived disk structure. 
\end{enumerate}

\acknowledgments
We are thankful to the anonymous referee for a careful and thorough report which improved the quality of this manuscript. MR and UG acknowledge support by the National Aeronautics and Space Administration through the NASA Astrobiology Institute under Cooperative Agreement Notice NNH13ZDA017C issued through the Science Mission Directorate. UG also acknowledges support from an award through NASA XRP 80NSSC20K0273.

\appendix
\section{Reduced network for CO ice chemistry and snowline location}
\label{sec:co_snowline_loc}
The location of the CO snowline can be obtained by solving the set of reactions:

\begin{eqnarray*}
{\rm H/H_2} &\rightarrow& {\rm sH/sH_2} ~ (k_\text{acc}) \\
{\rm sH/sH_2} &\rightarrow& {\rm H/H_2} ~ (k_\text{des}) \\
{\rm O} &\rightarrow& {\rm sO} ~ (k_\text{acc}) \\
{\rm sO + sH} &\rightarrow& {\rm sOH} ~ (k_\text{reac}) \\
{\rm sOH + sH/sH_2} &\rightarrow& {\rm sH_2O} ~ (k_\text{reac}) \\
{\rm sO} &\rightarrow& {\rm O} ~ (k_\text{des}) \\
{\rm sH_2O} + \gamma &\rightarrow& {\rm sOH + sH} ~ (k_\text{diss}) \\
{\rm CO} &\rightarrow& {\rm sCO} ~ (k_\text{acc}) \\
{\rm sCO} &\rightarrow& {\rm CO} ~ (k_\text{des}) \\
{\rm sCO + sOH} &\rightarrow& {\rm sCO_2 + sH} ~(k_\text{reac}; \Delta E = 150 \text{K}) \\
{\rm sCO_2} &\rightarrow& { \rm CO } + { \rm O }~ (k_\text{des}) 
\end{eqnarray*}

In this reduced network, we neglect the desorption of sOH and sH$_2$O from the grains and assume that CO$_2$ is instantly photodissociated back to CO upon desorption (last reaction). In practice, this simplified network does not allow an analytical solution and must be solved numerically.
In this set of reactions, $k_\text{acc}$ is the accretion rate and is given by

\begin{equation}
k_\text{acc}(i) = v_i \langle \sigma_d n_d \rangle
\end{equation}

where $v_i$ is the thermal velocity of gas species $i$ and $\langle \sigma_d n_d \rangle$ the product of the dust cross-sectional area and the dust density averaged over the grain size distribution. Note that this expression assumes that the sticking coefficient is equal to 1.
$k_\text{reac}$ is the reaction rate at the surface and is computed by

\begin{equation}
\label{eq:reac}
    k_\text{reac}(ij) = \kappa_{ij}\frac{k_{\text{hop}}(i) + k_{\text{hop}}(j)}{4 \langle \sigma_d n_d \rangle N_\text{sites}}
\end{equation}

where $\kappa_{ij}$ is the probability that the reaction occurs, $k_{\text{hop}}(i)$ the thermal hopping rate of species $i$ and $N_\text{sites}\sim 10^{15}$ cm$^{-2}$ the number of adsorption sites per cm$^2$. The thermal hopping rate is given by 

\begin{equation}\label{eq:thop}
k_\text{hop}(i) = \nu_i \exp\big(-E_\text{diff}(i)/T_\text{d}\big)
\end{equation}

where $T_d$ is the dust temperature, $ \nu_i=\sqrt{2N_\text{sites} k E_\text{bind}(i)/ \pi^2 m_i}$ is the vibrational frequency of the species $i$, $E_\text{bind}$ its binding energy, $m_i$ its mass and $E_\text{diff}(i)$ the barrier against diffusion. $\kappa_{ij}$ is equal to 1 for all reactions that don't have a barrier. For reactions with barrier $\kappa_{ij}$ is computed as the result of the competition among reaction and diffusion \citep[][]{Garrod11,Ruaud16}

\begin{equation}
    \kappa_{ij} = \frac{\max(\nu_{i},\nu_{j})\kappa_{ij,0}}{\max(\nu_{i},\nu_{j})\kappa_{ij,0} + k_{\text{hop},i} + k_{\text{hop},j} }
\end{equation}

where $\kappa_{ij,0}$ is expressed as the $\exp(-\Delta E_{ij}/ T_\text{d})$ with $\Delta E_{ij}$ the activation barrier of the reaction. For reactions involving H and H$_2$ we use the quantum mechanical probability for tunneling through a rectangular barrier of thickness $a=1$\AA; $\kappa_{ij,0}=\exp[-2(a/ \hbar)(2\mu \Delta E_{ij})^{1/2}]$, with $\mu$ the reduced mass of the system.
Finally, $k_\text{des}$ is the desorption rate by both thermal and non-thermal processes (i.e. photodesorption in this case)

\begin{equation}
k_\text{des}(i) = \Big[ \nu_i \exp\big(-E_\text{bind}(i)/T_\text{d}\big) + \frac{Y_\text{pd}(i) F_\text{FUV}}{4 N_\text{sites} N_\text{act}}\Big] \eta_\text{des}
\end{equation}

where $E_\text{bind}$ is the binding energy of species $i$, $Y_\text{pd}$ its photodesorption yield, $F_\text{FUV}$ is the incident FUV photon flux on the grain, $N_\text{act}$ the number of active surface layers (taken to be equal to 2 in our model) and $\eta_\text{des}$ the fraction of molecular species in the top desorbable surface layers and given by

\begin{equation}
\eta^\text{des} = 
\begin{cases}
    1,  & \text{if we have less than $N_\text{lay}$ monolayers of ice}\\ 
    \frac{4 \langle \sigma_d n_d \rangle N_\text{sites} N_\text{act}}{n_\text{ice}}, & \text{otherwise.}
\end{cases}
\end{equation}

Finally, $k_\text{diss}$ is the photodissociation rate by both direct FUV photons and CR generated photons and is given by

\begin{equation}
k_\text{diss}(i) = G_0 \alpha_i \exp\big(-\beta_i A_V\big) + \gamma_i \zeta_{\text{H}_2} f_{\text{H}_2}
\end{equation}

where $G_0$ is the integrated FUV flux with respect to that of the ISRF in units of the Draine field in the 6-13.6 eV range,  $\alpha$ is the unattenuated photorate, $\gamma$ the cosmic ray induced photodissociation efficiency, $\zeta_{\text{H}_2}$ the cosmic ray ionization rate of H$_2$ and $f_{\text{H}_2} = n(\text{H}_2)/[n(\text{H}) + 2n(\text{H}_2)]$ the molecular fraction. For water ice, we use $\alpha = 10^{-9}$ s$^{-1}$, $\beta = 1.8$ and $\gamma = 10^3$. Binding energies used for each of the species included in the reduced network are given in Table \ref{tab:iceparam}. For diffusion energy barriers we assume a fixed fraction of the binding energy of $0.4 \times E_\text{bind}$. For more details on the different parameters used here see \citet{Ruaud16} and \citet{Ruaud19}.

\begin{table}
    \centering
    \caption{Binding energies and diffusion energy barriers used for each of the species included in the reduced network.}
    \begin{tabular}{l c c}
    \hline
    \hline
    Species & Binding energy (K)& Reference\\
    \hline
    sH       &   650     &   1 \\
    sH$_2$   &   440     &   2 \\
    sO       &   800     &   3 \\
    sCO      &   1150    &   3 \\
    sCO$_2$  &   2575    &   3 \\
    sOH      &   2850    &   3 \\
    sH$_2$O  &   5700    &   3 \\
\hline
    \end{tabular}
    \tablerefs{$^{(a)}$\citet{AlHalabi07}, $^{(2)}$\citet{Amiaud07}, $^{(3)}$\citet{Garrod06}.}
    \label{tab:iceparam}
\end{table}

\bibliographystyle{aasjournal}
\bibliography{bibliography}

\begin{thebibliography}{}
\expandafter\ifx\csname natexlab\endcsname\relax\def\natexlab#1{#1}\fi

\bibitem[{{Aikawa} {et~al.}(2015){Aikawa}, {Furuya}, {Nomura}, \&
  {Qi}}]{Aikawa15}
{Aikawa}, Y., {Furuya}, K., {Nomura}, H., \& {Qi}, C. 2015, \apj, 807, 120

\bibitem[{{Al-Halabi} \& {van Dishoeck}(2007)}]{AlHalabi07}
{Al-Halabi}, A., \& {van Dishoeck}, E.~F. 2007, \mnras, 382, 1648

\bibitem[{{Alcal{\'a}} {et~al.}(2017){Alcal{\'a}}, {Manara}, {Natta}, {Frasca},
  {Testi}, {Nisini}, {Stelzer}, {Williams}, {Antoniucci}, {Biazzo}, {Covino},
  {Esposito}, {Getman}, \& {Rigliaco}}]{Alcala17}
{Alcal{\'a}}, J.~M., {Manara}, C.~F., {Natta}, A., {et~al.} 2017, \aap, 600,
  A20

\bibitem[{{Amiaud} {et~al.}(2007){Amiaud}, {Dulieu}, {Fillion}, {Momeni}, \&
  {Lemaire}}]{Amiaud07}
{Amiaud}, L., {Dulieu}, F., {Fillion}, J.~H., {Momeni}, A., \& {Lemaire}, J.~L.
  2007, \jcp, 127, 144709

\bibitem[{{Ansdell} {et~al.}(2017){Ansdell}, {Williams}, {Manara}, {Miotello},
  {Facchini}, {van der Marel}, {Testi}, \& {van Dishoeck}}]{Ansdell17}
{Ansdell}, M., {Williams}, J.~P., {Manara}, C.~F., {et~al.} 2017, \aj, 153, 240

\bibitem[{{Ansdell} {et~al.}(2016){Ansdell}, {Williams}, {van der Marel},
  {Carpenter}, {Guidi}, {Hogerheijde}, {Mathews}, {Manara}, {Miotello},
  {Natta}, {Oliveira}, {Tazzari}, {Testi}, {van Dishoeck}, \& {van
  Terwisga}}]{Ansdell16}
{Ansdell}, M., {Williams}, J.~P., {van der Marel}, N., {et~al.} 2016, \apj,
  828, 46

\bibitem[{{Ballering} \& {Eisner}(2019)}]{Ballering19}
{Ballering}, N.~P., \& {Eisner}, J.~A. 2019, \aj, 157, 144

\bibitem[{{Bergin} {et~al.}(2014){Bergin}, {Cleeves}, {Crockett}, \&
  {Blake}}]{Bergin14}
{Bergin}, E.~A., {Cleeves}, L.~I., {Crockett}, N., \& {Blake}, G.~A. 2014,
  Faraday Discussions, 168, 61

\bibitem[{{Bergin} {et~al.}(2016){Bergin}, {Du}, {Cleeves}, {Blake}, {Schwarz},
  {Visser}, \& {Zhang}}]{Bergin16}
{Bergin}, E.~A., {Du}, F., {Cleeves}, L.~I., {et~al.} 2016, \apj, 831, 101

\bibitem[{{Bergin} \& {Williams}(2017)}]{Bergin18}
{Bergin}, E.~A., \& {Williams}, J.~P. 2017, {The Determination of
  Protoplanetary Disk Masses}, ed. M.~{Pessah} \& O.~{Gressel}, Vol. 445, 1

\bibitem[{{Bergin} {et~al.}(2013){Bergin}, {Cleeves}, {Gorti}, {Zhang},
  {Blake}, {Green}, {Andrews}, {Evans}, {Henning}, {{\"O}berg}, {Pontoppidan},
  {Qi}, {Salyk}, \& {van Dishoeck}}]{Bergin13}
{Bergin}, E.~A., {Cleeves}, L.~I., {Gorti}, U., {et~al.} 2013, \nat, 493, 644

\bibitem[{{Bosman} {et~al.}(2018){Bosman}, {Walsh}, \& {van
  Dishoeck}}]{Bosman18}
{Bosman}, A.~D., {Walsh}, C., \& {van Dishoeck}, E.~F. 2018, \aap, 618, A182

\bibitem[{{Bruderer} {et~al.}(2012){Bruderer}, {van Dishoeck}, {Doty}, \&
  {Herczeg}}]{Bruderer12}
{Bruderer}, S., {van Dishoeck}, E.~F., {Doty}, S.~D., \& {Herczeg}, G.~J. 2012,
  \aap, 541, A91

\bibitem[{{Calahan} {et~al.}(2021{\natexlab{a}}){Calahan}, {Bergin}, {Zhang},
  {Schwarz}, {Oberg}, {Guzman}, {Walsh}, {Aikawa}, {Alarcon}, {Andrews}, {Bae},
  {Bergner}, {Booth}, {Bosman}, {Cataldi}, {Czekala}, {Huang}, {Ilee}, {Law},
  {Le Gal}, {Long}, {Loomis}, {Menard}, {Nomura}, {Qi}, {Teague}, {van'T Hoff},
  {Wilner}, \& {Yamato}}]{Calahanb}
{Calahan}, J.~K., {Bergin}, E.~A., {Zhang}, K., {et~al.} 2021{\natexlab{a}},
  arXiv e-prints, arXiv:2109.06202

\bibitem[{{Calahan} {et~al.}(2021{\natexlab{b}}){Calahan}, {Bergin}, {Zhang},
  {Teague}, {Cleeves}, {Bergner}, {Blake}, {Cazzoletti}, {Guzm{\'a}n},
  {Hogerheijde}, {Huang}, {Kama}, {Loomis}, {{\"O}berg}, {Qi}, {van Dishoeck},
  {Terwisscha van Scheltinga}, {Walsh}, \& {Wilner}}]{Calahan21}
{Calahan}, J.~K., {Bergin}, E., {Zhang}, K., {et~al.} 2021{\natexlab{b}}, \apj,
  908, 8

\bibitem[{{Calvet} {et~al.}(2002){Calvet}, {D'Alessio}, {Hartmann}, {Wilner},
  {Walsh}, \& {Sitko}}]{Calvet02}
{Calvet}, N., {D'Alessio}, P., {Hartmann}, L., {et~al.} 2002, \apj, 568, 1008

\bibitem[{{Calvet} {et~al.}(1991){Calvet}, {Patino}, {Magris}, \&
  {D'Alessio}}]{Calvet91}
{Calvet}, N., {Patino}, A., {Magris}, G.~C., \& {D'Alessio}, P. 1991, \apj,
  380, 617

\bibitem[{{Chiang} \& {Goldreich}(1997)}]{chiang97}
{Chiang}, E.~I., \& {Goldreich}, P. 1997, \apj, 490, 368

\bibitem[{{Cleeves} {et~al.}(2021){Cleeves}, {Loomis}, {Teague}, {Bergin},
  {Wilner}, {Bergner}, {Blake}, {Calahan}, {Cazzoletti}, {van Dishoeck},
  {Guzman}, {Hogerheijde}, {Huang}, {Kama}, {Oberg}, {Qi}, {Terwisscha van
  Scheltinga}, \& {Walsh}}]{Cleeves21}
{Cleeves}, L.~I., {Loomis}, R.~A., {Teague}, R., {et~al.} 2021, arXiv e-prints,
  arXiv:2102.09577

\bibitem[{{D'Alessio} {et~al.}(1998){D'Alessio}, {Cant{\"o}}, {Calvet}, \&
  {Lizano}}]{dalessio98}
{D'Alessio}, P., {Cant{\"o}}, J., {Calvet}, N., \& {Lizano}, S. 1998, \apj,
  500, 411

\bibitem[{{Draine} \& {Bertoldi}(1996)}]{Draine96}
{Draine}, B.~T., \& {Bertoldi}, F. 1996, \apj, 468, 269

\bibitem[{{Du} {et~al.}(2015){Du}, {Bergin}, \& {Hogerheijde}}]{Du15}
{Du}, F., {Bergin}, E.~A., \& {Hogerheijde}, M.~R. 2015, \apjl, 807, L32

\bibitem[{{Dullemond} {et~al.}(2001){Dullemond}, {Dominik}, \&
  {Natta}}]{Dullemond2001}
{Dullemond}, C.~P., {Dominik}, C., \& {Natta}, A. 2001, \apj, 560, 957

\bibitem[{{Dullemond} {et~al.}(2012){Dullemond}, {Juhasz}, {Pohl}, {Sereshti},
  {Shetty}, {Peters}, {Commercon}, \& {Flock}}]{Dullemond12}
{Dullemond}, C.~P., {Juhasz}, A., {Pohl}, A., {et~al.} 2012, {RADMC-3D: A
  multi-purpose radiative transfer tool}, , , ascl:1202.015

\bibitem[{{Eistrup} {et~al.}(2016){Eistrup}, {Walsh}, \& {van
  Dishoeck}}]{Eistrup16}
{Eistrup}, C., {Walsh}, C., \& {van Dishoeck}, E.~F. 2016, \aap, 595, A83

\bibitem[{{Favre} {et~al.}(2013){Favre}, {Cleeves}, {Bergin}, {Qi}, \&
  {Blake}}]{Favre13}
{Favre}, C., {Cleeves}, L.~I., {Bergin}, E.~A., {Qi}, C., \& {Blake}, G.~A.
  2013, \apjl, 776, L38

\bibitem[{{Furuya} \& {Aikawa}(2014)}]{Furuya14}
{Furuya}, K., \& {Aikawa}, Y. 2014, \apj, 790, 97

\bibitem[{{Garrod} \& {Herbst}(2006)}]{Garrod06}
{Garrod}, R.~T., \& {Herbst}, E. 2006, \aap, 457, 927

\bibitem[{{Garrod} \& {Pauly}(2011)}]{Garrod11}
{Garrod}, R.~T., \& {Pauly}, T. 2011, \apj, 735, 15

\bibitem[{{Gong} {et~al.}(2017){Gong}, {Ostriker}, \& {Wolfire}}]{Gong17}
{Gong}, M., {Ostriker}, E.~C., \& {Wolfire}, M.~G. 2017, \apj, 843, 38

\bibitem[{{Gorti} \& {Hollenbach}(2004)}]{Gorti04}
{Gorti}, U., \& {Hollenbach}, D. 2004, \apj, 613, 424

\bibitem[{{Gorti} \& {Hollenbach}(2008)}]{Gorti08}
---. 2008, \apj, 683, 287

\bibitem[{{Gorti} {et~al.}(2015){Gorti}, {Hollenbach}, \&
  {Dullemond}}]{Gorti15}
{Gorti}, U., {Hollenbach}, D., \& {Dullemond}, C.~P. 2015, \apj, 804, 29

\bibitem[{{Gorti} {et~al.}(2011){Gorti}, {Hollenbach}, {Najita}, \&
  {Pascucci}}]{Gorti11}
{Gorti}, U., {Hollenbach}, D., {Najita}, J., \& {Pascucci}, I. 2011, \apj, 735,
  90

\bibitem[{{Hildebrand}(1983)}]{Hildebrand1983}
{Hildebrand}, R.~H. 1983, \qjras, 24, 267

\bibitem[{{Hollenbach} \& {McKee}(1979)}]{Hollenbach79}
{Hollenbach}, D., \& {McKee}, C.~F. 1979, \apjs, 41, 555

\bibitem[{{Kama} {et~al.}(2016){Kama}, {Bruderer}, {Carney}, {Hogerheijde},
  {van Dishoeck}, {Fedele}, {Baryshev}, {Boland }, {G{\"u}sten}, {Aikutalp},
  {Choi}, {Endo}, {Frieswijk}, {Karska}, {Klaassen}, {Koumpia}, {Kristensen},
  {Leurini}, {Nagy}, {Perez Beaupuits}, {Risacher}, {van der Marel}, {van
  Kempen}, {van Weeren}, {Wyrowski}, \& {Y{\i}ld{\i}z}}]{Kama16}
{Kama}, M., {Bruderer}, S., {Carney}, M., {et~al.} 2016, \aap, 588, A108

\bibitem[{{Kama} {et~al.}(2020){Kama}, {Trapman}, {Fedele}, {Bruderer},
  {Hogerheijde}, {Miotello}, {van Dishoeck}, {Clarke}, \& {Bergin}}]{Kama20}
{Kama}, M., {Trapman}, L., {Fedele}, D., {et~al.} 2020, \aap, 634, A88

\bibitem[{{Krijt} {et~al.}(2020){Krijt}, {Bosman}, {Zhang}, {Schwarz},
  {Ciesla}, \& {Bergin}}]{Krijt20}
{Krijt}, S., {Bosman}, A.~D., {Zhang}, K., {et~al.} 2020, \apj, 899, 134

\bibitem[{{Krijt} {et~al.}(2016){Krijt}, {Ciesla}, \& {Bergin}}]{Krijt16}
{Krijt}, S., {Ciesla}, F.~J., \& {Bergin}, E.~A. 2016, \apj, 833, 285

\bibitem[{{Krijt} {et~al.}(2018){Krijt}, {Schwarz}, {Bergin}, \&
  {Ciesla}}]{Krijt18}
{Krijt}, S., {Schwarz}, K.~R., {Bergin}, E.~A., \& {Ciesla}, F.~J. 2018, \apj,
  864, 78

\bibitem[{{Loison} {et~al.}(2014){Loison}, {Wakelam}, {Hickson}, {Bergeat}, \&
  {Mereau}}]{Loison14}
{Loison}, J.-C., {Wakelam}, V., {Hickson}, K.~M., {Bergeat}, A., \& {Mereau},
  R. 2014, \mnras, 437, 930

\bibitem[{{Loison} {et~al.}(2019){Loison}, {Wakelam}, {Gratier}, {Hickson},
  {Bacmann}, {Ag{\'u}ndez}, {Marcelino}, {Cernicharo}, {Guzman}, {Gerin},
  {Goicoechea}, {Roueff}, {Petit}, {Pety}, {Fuente}, \&
  {Riviere-Marichalar}}]{Loison19}
{Loison}, J.-C., {Wakelam}, V., {Gratier}, P., {et~al.} 2019, \mnras, 485, 5777

\bibitem[{{Long} {et~al.}(2017){Long}, {Herczeg}, {Pascucci}, {Drabek-Maunder},
  {Mohanty}, {Testi}, {Apai}, {Hendler}, {Henning}, {Manara}, \&
  {Mulders}}]{Long17}
{Long}, F., {Herczeg}, G.~J., {Pascucci}, I., {et~al.} 2017, \apj, 844, 99

\bibitem[{{Mac{\'\i}as} {et~al.}(2021){Mac{\'\i}as}, {Guerra-Alvarado},
  {Carrasco-Gonz{\'a}lez}, {Ribas}, {Espaillat}, {Huang}, \&
  {Andrews}}]{Macias21}
{Mac{\'\i}as}, E., {Guerra-Alvarado}, O., {Carrasco-Gonz{\'a}lez}, C., {et~al.}
  2021, \aap, 648, A33

\bibitem[{{Manara} {et~al.}(2016){Manara}, {Rosotti}, {Testi}, {Natta},
  {Alcal{\'a}}, {Williams}, {Ansdell}, {Miotello}, {van der Marel}, {Tazzari},
  {Carpenter}, {Guidi}, {Mathews}, {Oliveira}, {Prusti}, \& {van
  Dishoeck}}]{Manara16}
{Manara}, C.~F., {Rosotti}, G., {Testi}, L., {et~al.} 2016, \aap, 591, L3

\bibitem[{{McClure}(2019)}]{McClure19}
{McClure}, M.~K. 2019, \aap, 632, A32

\bibitem[{{McClure} {et~al.}(2016){McClure}, {Bergin}, {Cleeves}, {van
  Dishoeck}, {Blake}, {Evans}, {Green}, {Henning}, {{\"O}berg}, {Pontoppidan},
  \& {Salyk}}]{Mcclure2016}
{McClure}, M.~K., {Bergin}, E.~A., {Cleeves}, L.~I., {et~al.} 2016, \apj, 831,
  167

\bibitem[{{Miotello} {et~al.}(2014){Miotello}, {Bruderer}, \& {van
  Dishoeck}}]{Miotello14}
{Miotello}, A., {Bruderer}, S., \& {van Dishoeck}, E.~F. 2014, \aap, 572, A96

\bibitem[{{Miotello} {et~al.}(2016){Miotello}, {van Dishoeck}, {Kama}, \&
  {Bruderer}}]{Miotello16}
{Miotello}, A., {van Dishoeck}, E.~F., {Kama}, M., \& {Bruderer}, S. 2016,
  \aap, 594, A85

\bibitem[{{Miotello} {et~al.}(2017){Miotello}, {van Dishoeck}, {Williams},
  {Ansdell}, {Guidi}, {Hogerheijde}, {Manara}, {Tazzari}, {Testi}, {van der
  Marel}, \& {van Terwisga}}]{Miotello17}
{Miotello}, A., {van Dishoeck}, E.~F., {Williams}, J.~P., {et~al.} 2017, \aap,
  599, A113

\bibitem[{{Molyarova} {et~al.}(2017{\natexlab{a}}){Molyarova}, {Akimkin},
  {Semenov}, {Henning}, {Vasyunin}, \& {Wiebe}}]{Molyarova17}
{Molyarova}, T., {Akimkin}, V., {Semenov}, D., {et~al.} 2017{\natexlab{a}},
  \apj, 849, 130

\bibitem[{{Molyarova} {et~al.}(2017{\natexlab{b}}){Molyarova}, {Akimkin},
  {Semenov}, {Henning}, {Vasyunin}, \& {Wiebe}}]{Molyarova2017}
---. 2017{\natexlab{b}}, \apj, 849, 130

\bibitem[{{{\"O}berg} \& {Bergin}(2021)}]{Oberg21}
{{\"O}berg}, K.~I., \& {Bergin}, E.~A. 2021, \physrep, 893, 1

\bibitem[{{Oberg} {et~al.}(2021){Oberg}, {Guzman}, {Walsh}, {Aikawa}, {Bergin},
  {Law}, {Loomis}, {Alarcon}, {Andrews}, {Bae}, {Bergner}, {Boehler}, {Booth},
  {Bosman}, {Calahan}, {Cataldi}, {Cleeves}, {Czekala}, {Furuya}, {Huang},
  {Ilee}, {Kurtovic}, {Le Gal}, {Liu}, {Long}, {Menard}, {Nomura}, {Perez},
  {Qi}, {Schwarz}, {Sierra}, {Teague}, {Tsukagoshi}, {Yamato}, {van 't Hoff},
  {Waggoner}, {Wilner}, \& {Zhang}}]{Oberg21b}
{Oberg}, K.~I., {Guzman}, V.~V., {Walsh}, C., {et~al.} 2021, arXiv e-prints,
  arXiv:2109.06268

\bibitem[{{Padovani} {et~al.}(2018){Padovani}, {Ivlev}, {Galli}, \&
  {Caselli}}]{Padovani18}
{Padovani}, M., {Ivlev}, A.~V., {Galli}, D., \& {Caselli}, P. 2018, \aap, 614,
  A111

\bibitem[{{Pascucci} {et~al.}(2016){Pascucci}, {Testi}, {Herczeg}, {Long},
  {Manara}, {Hendler}, {Mulders}, {Krijt}, {Ciesla}, {Henning}, {Mohanty},
  {Drabek-Maunder}, {Apai}, {Sz{\H{u}}cs}, {Sacco}, \& {Olofsson}}]{Pascucci16}
{Pascucci}, I., {Testi}, L., {Herczeg}, G.~J., {et~al.} 2016, \apj, 831, 125

\bibitem[{{Powell} {et~al.}(2019){Powell}, {Murray-Clay}, {P{\'e}rez},
  {Schlichting}, \& {Rosenthal}}]{Powell19}
{Powell}, D., {Murray-Clay}, R., {P{\'e}rez}, L.~M., {Schlichting}, H.~E., \&
  {Rosenthal}, M. 2019, \apj, 878, 116

\bibitem[{{Reboussin} {et~al.}(2015){Reboussin}, {Wakelam}, {Guilloteau},
  {Hersant}, \& {Dutrey}}]{Reboussin15}
{Reboussin}, L., {Wakelam}, V., {Guilloteau}, S., {Hersant}, F., \& {Dutrey},
  A. 2015, \aap, 579, A82

\bibitem[{{Ribas} {et~al.}(2020){Ribas}, {Espaillat}, {Mac{\'\i}as}, \&
  {Sarro}}]{Ribas20}
{Ribas}, {\'A}., {Espaillat}, C.~C., {Mac{\'\i}as}, E., \& {Sarro}, L.~M. 2020,
  \aap, 642, A171

\bibitem[{{Roueff} {et~al.}(2015){Roueff}, {Loison}, \& {Hickson}}]{Roueff15}
{Roueff}, E., {Loison}, J.~C., \& {Hickson}, K.~M. 2015, \aap, 576, A99

\bibitem[{{Ruaud} \& {Gorti}(2019)}]{Ruaud19}
{Ruaud}, M., \& {Gorti}, U. 2019, \apj, 885, 146

\bibitem[{{Ruaud} {et~al.}(2016){Ruaud}, {Wakelam}, \& {Hersant}}]{Ruaud16}
{Ruaud}, M., {Wakelam}, V., \& {Hersant}, F. 2016, \mnras, 459, 3756

\bibitem[{{Schwarz} {et~al.}(2016){Schwarz}, {Bergin}, {Cleeves}, {Blake},
  {Zhang}, {{\"O}berg}, {van Dishoeck}, \& {Qi}}]{Schwarz16}
{Schwarz}, K.~R., {Bergin}, E.~A., {Cleeves}, L.~I., {et~al.} 2016, \apj, 823,
  91

\bibitem[{{Schwarz} {et~al.}(2021){Schwarz}, {Calahan}, {Zhang}, {Alarc{\'o}n},
  {Aikawa}, {Andrews}, {Bae}, {Bergin}, {Booth}, {Bosman}, {Cataldi},
  {Cleeves}, {Czekala}, {Huang}, {Ilee}, {Law}, {Le Gal}, {Liu}, {Long},
  {Loomis}, {Mac{\'\i}as}, {McClure}, {M{\'e}nard}, {{\"O}berg}, {Teague}, {van
  Dishoeck}, {Walsh}, \& {Wilner}}]{Schwarz21}
{Schwarz}, K.~R., {Calahan}, J.~K., {Zhang}, K., {et~al.} 2021, arXiv e-prints,
  arXiv:2109.06228

\bibitem[{{Teague} {et~al.}(2019){Teague}, {Bae}, {Huang}, \&
  {Bergin}}]{Teague19}
{Teague}, R., {Bae}, J., {Huang}, J., \& {Bergin}, E.~A. 2019, \apjl, 884, L56

\bibitem[{{Tielens} \& {Hollenbach}(1985)}]{Tielens85}
{Tielens}, A.~G.~G.~M., \& {Hollenbach}, D. 1985, \apj, 291, 722

\bibitem[{{Trapman} {et~al.}(2021){Trapman}, {Bosman}, {Rosotti},
  {Hogerheijde}, \& {van Dishoeck}}]{Trapman2021}
{Trapman}, L., {Bosman}, A.~D., {Rosotti}, G., {Hogerheijde}, M.~R., \& {van
  Dishoeck}, E.~F. 2021, arXiv e-prints, arXiv:2103.05654

\bibitem[{{Trapman} {et~al.}(2017){Trapman}, {Miotello}, {Kama}, {van
  Dishoeck}, \& {Bruderer}}]{Trapman17}
{Trapman}, L., {Miotello}, A., {Kama}, M., {van Dishoeck}, E.~F., \&
  {Bruderer}, S. 2017, \aap, 605, A69

\bibitem[{{Tychoniec} {et~al.}(2020){Tychoniec}, {Manara}, {Rosotti}, {van
  Dishoeck}, {Cridland}, {Hsieh}, {Murillo}, {Segura-Cox}, {van Terwisga}, \&
  {Tobin}}]{Tychoniec2020}
{Tychoniec}, {\L}., {Manara}, C.~F., {Rosotti}, G.~P., {et~al.} 2020, \aap,
  640, A19

\bibitem[{{Visser} {et~al.}(2009){Visser}, {van Dishoeck}, \&
  {Black}}]{Visser09}
{Visser}, R., {van Dishoeck}, E.~F., \& {Black}, J.~H. 2009, \aap, 503, 323

\bibitem[{{Williams} \& {Best}(2014)}]{Williams14}
{Williams}, J.~P., \& {Best}, W. M.~J. 2014, \apj, 788, 59

\bibitem[{{Woitke} {et~al.}(2016){Woitke}, {Min}, {Pinte}, {Thi}, {Kamp},
  {Rab}, {Anthonioz}, {Antonellini}, {Baldovin-Saavedra}, {Carmona}, {Dominik},
  {Dionatos}, {Greaves}, {G{\"u}del}, {Ilee}, {Liebhart}, {M{\'e}nard},
  {Rigon}, {Waters}, {Aresu}, {Meijerink}, \& {Spaans}}]{Woitke2016}
{Woitke}, P., {Min}, M., {Pinte}, C., {et~al.} 2016, \aap, 586, A103

\bibitem[{{Xu} {et~al.}(2017){Xu}, {Bai}, \& {{\"O}berg}}]{Xu17}
{Xu}, R., {Bai}, X.-N., \& {{\"O}berg}, K. 2017, \apj, 835, 162

\bibitem[{{Zhang} {et~al.}(2017){Zhang}, {Bergin}, {Blake}, {Cleeves}, \&
  {Schwarz}}]{Zhang17}
{Zhang}, K., {Bergin}, E.~A., {Blake}, G.~A., {Cleeves}, L.~I., \& {Schwarz},
  K.~R. 2017, Nature Astronomy, 1, 0130

\bibitem[{{Zhang} {et~al.}(2019){Zhang}, {Bergin}, {Schwarz}, {Krijt}, \&
  {Ciesla}}]{Zhang19}
{Zhang}, K., {Bergin}, E.~A., {Schwarz}, K., {Krijt}, S., \& {Ciesla}, F. 2019,
  \apj, 883, 98

\bibitem[{{Zhang} {et~al.}(2021){Zhang}, {Booth}, {Law}, {Bosman}, {Schwarz},
  {Bergin}, {{\"O}berg}, {Andrews}, {Guzm{\'a}n}, {Walsh}, {Qi}, {van 't Hoff},
  {Long}, {Wilner}, {Huang}, {Czekala}, {Ilee}, {Cataldi}, {Bergner}, {Aikawa},
  {Teague}, {Bae}, {Loomis}, {Calahan}, {Alarc{\'o}n}, {M{\'e}nard}, {Le Gal},
  {Sierra}, {Yamato}, {Nomura}, {Tsukagoshi}, {P{\'e}rez}, {Trapman}, {Liu}, \&
  {Furuya}}]{Zhang21}
{Zhang}, K., {Booth}, A.~S., {Law}, C.~J., {et~al.} 2021, arXiv e-prints,
  arXiv:2109.06233

\end{thebibliography}

\end{document}